\documentclass{article}
\begin{document}
{\LARGE
\begin{center}
{\bf
Charmed ${\bf (70,1^-)}$ baryon multiplet}
\end{center}
}
\vskip3ex
\noindent
S.M. Gerasyuta and E.E. Matskevich

\vskip2ex
\noindent
Department of Theoretical Physics, St. Petersburg State University,
198904, St. Petersburg, Russia

\noindent
Department of Physics, LTA, 194021, St. Petersburg, Russia

\vskip4ex
\begin{center}
{\bf Abstract}
\end{center}
\vskip4ex
{\large
The masses of negative parity $(70,1^-)$ charmed nonstrange baryons
are calculated in the relativistic quark model. The relativistic
three-quark equations of the $(70,1^-)$ charmed baryon multiplet
are found in the framework of the dispersion relation technique.
The approximate solutions of these equations using the method based
on the extraction of leading singularities of the amplitude are
obtained. The calculated mass values of the $(70,1^-)$ charmed
baryons are in good agreement with the experimental data.
\vskip2ex
\noindent
e-mail address: gerasyuta@SG6488.spb.edu

\noindent
e-mail address: matskev@pobox.spbu.ru
\vskip2ex
\noindent
PACS: 11.55.Fv, 12.39.Ki, 12.40.Yx, 14.20.Lq.
\vskip2ex
{\bf 1. Introduction.}
\vskip2ex
For many years CLEO was the main source of data on orbitally-excited
charmed baryons [1]. An excited $\Sigma_c$ candidate has been seen decaying
to $\Lambda_c \, \pi^+$, with mass about $510\, MeV$ above
$M(\Lambda_c)$ [2]. The first excitation of the $\Lambda_c$ and $\Xi_c$
scale well from the first $\Lambda$ excitations
$\Lambda (1405,\frac{1}{2}^-)$ and $\Lambda (1520,\frac{3}{2}^-)$. The
highest $\Lambda_c$ was seen by BaBar in decay mode $D^0 p$ [3]. The
highest $\Xi_c$ were reported by the Belle Collaboration in Ref. [4] and
confirmed by BaBar [5].

In the recent reviews [6, 7] the spectroscopy of hadrons containing heavy
quarks and some of their theoretical interpretation are given. One discuss
progress on orbitally excited charmed baryons.

In the series of papers [8 -- 12] a practical treatment of relativistic
three-hadron systems have been developed. The physics of three-hadron
system is usefully described in term of pairwise interactions among the
three particles. The theory is based on the two principles of unitarity
and analyticity, as applied to the two-body subenergy channels. The linear
integral equation in a single variable are obtained for the isobar
amplitudes.

Instead of the quadrature methods of obtaining solution the set of suitable
functions is identified and used as basis set for the expansion of the
desired solutions. By this means the couple integral equation are solved
in terms of simple algebra.

In our papers [13, 14] relativistic generalization of the three-body
Faddeev equations was obtained in the form of dispersion relations in the
pair energy of two interacting particles. The mass spectrum of $S$-wave
baryons including $u$, $d$, $s$-quarks was calculated by a method based
on isolating the leading singularities in the amplitude. We searched for
the approximate solution of integral three-quark equations by taking into
account two-particle and triangle singularities, all the weaker ones being
neglected. If we considered such an approximation, which corresponds to
taking into account two-body and triangle singularities, defined all the
smooth functions of the subenergy variables (as compared with the singular
part of the amplitude) in the middle point of the physical region of
Dalitz-plot, then the problem was reduced to the one of solving a system
of simple algebraic equations.

In the recent paper [15] the relativistic three-quark equations of the
excited $(70,1^-)$ baryons are found in the framework of the dispersion
relation technique. In our paper the orbital-spin-flavor wave functions
are constructed. We have used the orbital-spin-flavor wave functions for
the construction of integral equations. We take into account the $u$, $d$,
$s$-quarks. We have represented the $30$ nonstrange and strange resonances
belonging to the $(70,1^-)$ multiplet. The $15$ resonances are in good
agreement with experimental data. We have predicted $15$ masses of baryons.
In our model the four parameters are used: gluon coupling constants $g_+$
and $g_-$ for the various parity, cutoff energy parameters $\lambda$,
$\lambda_s$ for the nonstrange and strange diquarks.

The present paper is organized as follows. Section 2 is devoted to the
construction of the orbital-spin-flavor wave functions for the charmed
baryons $(70,1^-)$ multiplet. In Section 3 the relativistic three-quark
equations are constructed in the form of the dispersion relation over
the two-body subenergy. In Section 4 the systems of equations for the
reduced amplitudes are derived. Section 5 is devoted to the calculation
results for the $P$-wave charmed baryons mass spectrum (Tables I-IV).
In Conclusion, the status of the considered model is discussed.

\vskip2ex
{\bf 2. The wave function of ${\bf (70,1^-)}$ excited charmed states.}
\vskip2ex
Here we deal with a three-quark system having one unit of orbital
excitation. We take into account $u$, $d$, $c$-quarks. The orbital part of
wave function must have a mixed symmetry. The spin-flavor part of wave
function possesses the same symmetry in order to obtain a totally
symmetric state in the orbital-spin-flavor space.

For the sake of simplicity we derived the wave functions for the decuplets
$(10,2)$. The fully symmetric wave function for the decuplet state can be
constructed as [16].

$$\varphi=\frac{1}{\sqrt{2}}\left(
\varphi_{MA}^{SU(6)}\varphi_{MA}^{O(3)}+
\varphi_{MS}^{SU(6)}\varphi_{MS}^{O(3)}
\right).\eqno (1)$$

Then we obtain:

$$\varphi=\frac{1}{\sqrt{2}}\varphi_{S}^{SU(3)}\left(
\varphi_{MA}^{SU(2)}\varphi_{MA}^{O(3)}+
\varphi_{MS}^{SU(2)}\varphi_{MS}^{O(3)}
\right),\eqno (2)$$

\noindent
here $MA$ and $MS$ define the mixed antisymmetric and symmetric part
of wave function,

$$\varphi_{MA}^{SU(6)}=\varphi_{S}^{SU(3)}\varphi_{MA}^{SU(2)},\quad\quad
\varphi_{MS}^{SU(6)}=\varphi_{S}^{SU(3)}\varphi_{MS}^{SU(2)}.\eqno (3)$$

The functions $\varphi_{MA}^{SU(2)}$, $\varphi_{MS}^{SU(2)}$,
$\varphi_{MA}^{O(3)}$, $\varphi_{MS}^{O(3)}$ are given:

$$\varphi_{MA}^{SU(2)}=\frac{1}{\sqrt{2}}\left(
\uparrow \downarrow \uparrow-\downarrow \uparrow \uparrow
\right),\quad\quad
\varphi_{MS}^{SU(2)}=\frac{1}{\sqrt{6}}\left(
\uparrow \downarrow \uparrow+\downarrow \uparrow \uparrow-
2\uparrow \uparrow \downarrow
\right),\eqno (4)$$

$$\varphi_{MA}^{O(3)}=\frac{1}{\sqrt{2}}\left(
010-100
\right),\quad\quad
\varphi_{MS}^{O(3)}=\frac{1}{\sqrt{6}}\left(
010+100-2\cdot 001
\right).\eqno (5)$$

$\uparrow$ and $\downarrow$ determine the spin directions. $1$ and $0$
correspond to the excited or nonexcited quarks. The three projections
of orbital angular momentum are $l_z=1, 0, -1$. The $(10,2)$ multiplet with
$J^p=\frac{3}{2} ^{-}$ can be obtained using the spin $S=\frac{1}{2}$
and $l_z=1$, but the $(10,2)$ multiplet with $J^p=\frac{1}{2} ^{-}$
is determined by the spin $S=\frac{1}{2}$ and $l_z=0$.

We construct the $SU(3)$-function for each particle of multiplet.
For instance, the $SU(3)$-function for $\Sigma^{+}_c$-hyperon of decuplet
have following form:

$$\varphi_{S}^{SU(3)}=\frac{1}{\sqrt{3}}\left(
ucu+cuu+uuc
\right).\eqno (6)$$

We obtain the $SU(6)\times O(3)$-function for the $\Sigma^{+}_c$ of
the $(10,2)$ multiplet:

$$\varphi_{\Sigma^{+}_c(10,2)}=\frac{\sqrt{6}}{18}\left(
2\{u^1\downarrow u\uparrow c\uparrow\}+
\{c^1\downarrow u\uparrow u\uparrow\}-\right.$$
$$\left.-\{u^1\uparrow u\downarrow c\uparrow\}-
\{u^1\uparrow u\uparrow c\downarrow \}-
\{c^1\uparrow u\uparrow u\downarrow \}\right).\eqno (7)$$

Here the parenthesis determine the symmetrical function:

$$\{{abc}\}\equiv abc+acb+bac+cab+bca+cba.\eqno (8)$$

The wave functions of $\Sigma^{0}_c$- and $\Sigma^{-}_c$-hyperons can be
constructed by similar way.

For the $\Xi^{0,-}_{cc}$ state of the $(10,2)$ multiplet the wave function
is similar to the $\Sigma^{+,-}_c$ state with the replacement by
$u\leftrightarrow c$ or $d\leftrightarrow c$. The wave function for the
$\Omega_{ccc}$ of $(10,2)$ decuplet is determined as:

$$\varphi_{\Omega_{ccc}(10,2)}=\frac{\sqrt{2}}{6}\left(
\{c^1\downarrow c\uparrow c\uparrow\}
-\{c^1\uparrow c\uparrow c\downarrow\}
\right).\eqno (9)$$

The wave functions and the method of the construction for the multiplets
$(8,2)$, $(8,4)$ and $(1,2)$ are similar.

\vskip2ex
{\bf 3. The three-quark integral equations for the ${\bf (70,1^-)}$
multiplet.}
\vskip2ex
By the construction of $(70,1^-)$ charmed baryon multiplet integral
equations we need to using the projectors for the different diquark states.
The projectors to the symmetric and antisymmetric states can be obtained as:

$$\frac{1}{2}\left(q_1 q_2+q_2 q_1\right),\quad\quad
\frac{1}{2}\left(q_1 q_2-q_2 q_1\right).\eqno (10)$$

$$\frac{1}{2}\left(\uparrow\downarrow+\downarrow\uparrow
\right),\quad\quad
\frac{1}{2}\left(\uparrow\downarrow-\downarrow\uparrow
\right).\eqno (11)$$

$$\frac{1}{2}\left(10+01\right),\quad\quad
\frac{1}{2}\left(10-01\right).\eqno (12)$$

One can obtain the four types of totally symmetric projectors:

$$S=S\cdot S\cdot S=\frac{1}{8}\left(q_1 q_2+q_2 q_1\right)
\left(\uparrow\downarrow+\downarrow\uparrow\right)\left(10+01\right),
\eqno (13)$$

$$S=S\cdot A\cdot A=\frac{1}{8}\left(q_1 q_2+q_2 q_1\right)
\left(\uparrow\downarrow-\downarrow\uparrow\right)\left(10-01\right),
\eqno (14)$$

$$S=A\cdot A\cdot S=\frac{1}{8}\left(q_1 q_2-q_2 q_1\right)
\left(\uparrow\downarrow-\downarrow\uparrow\right)\left(10+01\right),
\eqno (15)$$

$$S=A\cdot S\cdot A=\frac{1}{8}\left(q_1 q_2-q_2 q_1\right)
\left(\uparrow\downarrow+\downarrow\uparrow\right)\left(10-01\right).
\eqno (16)$$

We use these projectors for the consideration of various diquarks:

\vskip2ex
\noindent
$u^1\uparrow c\downarrow\,\,:$

$$\frac{A^{0c}_{1}}{8}
\left(u^1\uparrow c\downarrow+u^1\downarrow c\uparrow+
c^1\uparrow u\downarrow+c^1\downarrow u\uparrow+
u\uparrow c^1\downarrow+u\downarrow c^1\uparrow+
c\uparrow u^1\downarrow+c\downarrow u^1\uparrow \right)$$
$$+\frac{A^{1c}_{0}}{8}
\left(u^1\uparrow c\downarrow-u^1\downarrow c\uparrow+
c^1\uparrow u\downarrow-c^1\downarrow u\uparrow-
u\uparrow c^1\downarrow+u\downarrow c^1\uparrow-
c\uparrow u^1\downarrow+c\downarrow u^1\uparrow \right)$$
$$+\frac{A^{0c}_{0}}{8}
\left(u^1\uparrow c\downarrow-u^1\downarrow c\uparrow-
c^1\uparrow u\downarrow+c^1\downarrow u\uparrow+
u\uparrow c^1\downarrow-u\downarrow c^1\uparrow-
c\uparrow u^1\downarrow+c\downarrow u^1\uparrow \right)$$
$$+\frac{A^{1c}_{1}}{8}
\left(u^1\uparrow c\downarrow+u^1\downarrow c\uparrow-
c^1\uparrow u\downarrow-c^1\downarrow u\uparrow-
u\uparrow c^1\downarrow-u\downarrow c^1\uparrow+
c\uparrow u^1\downarrow+c\downarrow u^1\uparrow \right),
\eqno (17)$$

\noindent
$u^1\uparrow c\uparrow\,\, :$

$$\frac{A^{0c}_{1}}{4}
\left(u^1\uparrow c\uparrow+c^1\uparrow u\uparrow+
u\uparrow c^1\uparrow+c\uparrow u^1\uparrow\right)+$$

$$+\frac{A^{1c}_{1}}{4}
\left(u^1\uparrow c\uparrow-c^1\uparrow u\uparrow-
u\uparrow c^1\uparrow+c\uparrow u^1\uparrow
\right),\eqno (18)$$

\noindent
$u\uparrow c\downarrow\,\, :$

$$\frac{A^{0c}_{1}}{4}
\left(u\uparrow c\downarrow+u\downarrow c\uparrow+
c\uparrow u\downarrow+c\downarrow u\uparrow
\right)+$$

$$+\frac{A^{0c}_{0}}{4}
\left(u\uparrow c\downarrow-u\downarrow c\uparrow-
c\uparrow u\downarrow+c\downarrow u\uparrow
\right),\eqno (19)$$

\noindent
$u\uparrow c\uparrow\,\, :$

$$\frac{A^{0c}_{1}}{2}
\left(u\uparrow c\uparrow+c\uparrow u\uparrow
\right).\eqno (20)$$

Here the lower index determines the value of spin projection,
and the upper index corresponds to the value of orbital angular momentum.

We consider the projectors (21)-(24), which are similar to (17)-(20) with
the replacement by $c\rightarrow u$ and use the amplitudes
$A^{0}_{1}$, $A^{1}_{0}$, $A^{0}_{0}$, $A^{1}_{1}$. The $A$ is the
three-quark amplitude.

\vskip2ex
\noindent
$u^1\uparrow u\downarrow\,\, :$

$$\frac{A^{0}_{1}}{4}
\left(u^1\uparrow u\downarrow+u^1\downarrow u\uparrow+
u\uparrow u^1\downarrow+u\downarrow u^1\uparrow
\right)+$$

$$+\frac{A^{1}_{0}}{4}
\left(u^1\uparrow u\downarrow-u^1\downarrow u\uparrow-
u\uparrow u^1\downarrow+u\downarrow u^1\uparrow
\right),\eqno (21)$$

\noindent
$u^1\uparrow u\uparrow\,\, :$

$$\frac{A^{0}_{1}}{2}
\left(u^1\uparrow u\uparrow+u\uparrow u^1\uparrow
\right),\eqno (22)$$

\noindent
$u\uparrow u\downarrow\,\, :$

$$\frac{A^{0}_{1}}{2}
\left(u\uparrow u\downarrow+u\downarrow u\uparrow
\right),\eqno (23)$$

\noindent
$u\uparrow u\uparrow\,\, :$

$$A^{0}_{1}\, u\uparrow u\uparrow .\eqno (24)$$

Here we consider the projection of orbital angular momentum $l_z=+1$. We
use only diquarks $1^+$, $0^+$, $2^-$, $1^-$. If we consider the $l_z=-1$
or $l_z=0$, that we obtain the other diquarks: $1^+$, $0^+$, $1^-$, $0^-$.
In our model the five types of diquarks $1^+$, $0^+$, $2^-$, $1^-$, $0^-$
are constructed.

For the sake of simplicity we derive the relativistic Faddeev equations
using the $\Sigma_c$ hyperon with $J^p=\frac{3}{2} ^{-}$ of the (10,2)
multiplets. We use the graphic equations for the amplitudes $A_J(s,s_{ik})$.
In order to represent the amplitude $A_J(s,s_{ik})$ in the form
of dispersion relations, it is necessary to define the amplitudes of
quark-quark interaction $a_J(s_{ik})$. The pair quarks amplitudes
$qq\rightarrow qq$ are calculated in the framework of the dispersion
$N/D$ method with the input four-fermion interaction with quantum numbers
of the gluon [17]. We use results of our relativistic quark model [18]
and write down the pair quark amplitudes in the form:

$$a_J(s_{ik})=\frac{G^2_J(s_{ik})}
{1-B_J(s_{ik})},\eqno (25)$$

$$B_J(s_{ik})=\int\limits_{(m_i+m_k)^2}^{\infty}\hskip2mm
\frac{ds'_{ik}}{\pi}\frac{\rho_J(s'_{ik})G^2_J(s'_{ik})}
{s'_{ik}-s_{ik}},\eqno (26)$$

$$\rho_J (s_{ik})=\frac{(m_i+m_k)^2}{4\pi}
\left(\alpha_J\frac{s_{ik}}{(m_i+m_k)^2}
+\beta_J +\frac{\delta_J}{s_{ik}} \right)\times$$

$$\times\frac{\sqrt{(s_{ik}-(m_i+m_k)^2)(s_{ik}-(m_i-m_k)^2)}}
{s_{ik}}\, .\eqno (27)$$

Here $G_J$ is the diquark vertex function; $B_J(s_{ik})$, $\rho_J (s_{ik})$
are the Chew-Mandelstam function [19] and the phase space consequently.
$s_{ik}$ is the two-particle subenergy squared (i,k=1,2,3), $s$ is the
systems total energy squared. For the state $J^p=\frac{3}{2} ^{-}$ of
the (10,2) multiplet we use three diquarks $J^p=1^+$, $1^+_c$, $1^-_c$.
The coefficients of Chew-Mandelstam function $\alpha_J$, $\beta_J$ and
$\delta_J$ are given in Table V.

In the case in question the interacting quarks do not produce bound state,
then the integration in dispersion integrals is carried out from
$(m_i+m_k)^2$ to $\infty$.

All diagrams are classified over the last quark pair (Fig.1).

We use the diquark projectors. We consider the particle $\Sigma_c$
$\frac{3}{2} ^{-}$ of the $(10,2)$ multiplet again. This wave function
contains the contribution $u^1\downarrow u\uparrow c\uparrow$, which
includes three diquarks: $u^1\downarrow u\uparrow$,\,
$u^1\downarrow c\uparrow$\, and \, $u\uparrow c\uparrow$.
The diquark projectors allow us to obtain the equations (28)-(30)
(with the definition $\rho_J(s_{ij})\equiv k_{ij}$).

$$k_{12}\left(\frac{A_1^0+A_0^1}{4}\left(u^1\downarrow u\uparrow c\uparrow
+u\uparrow u^1\downarrow c\uparrow\right)+\right.$$

$$\left. +\frac{A_1^0-A_0^1}{4}\left(
u^1\uparrow u\downarrow c\uparrow+u\downarrow u^1\uparrow c\uparrow
\right)\right)\, ,\eqno (28)$$

$$k_{13}\left(\frac{A_1^{0c}+A_0^{1c}+A_0^{0c}+A_1^{1c}}{8}
\left(u^1\downarrow u\uparrow c\uparrow
+c\uparrow u\uparrow u^1\downarrow\right)+\right.$$

$$\left. +\frac{A_1^{0c}-A_0^{1c}-A_0^{0c}+A_1^{1c}}{8}
\left(u^1\uparrow u\uparrow c\downarrow
+c\downarrow u\uparrow u^1\uparrow\right)+\right.$$

$$\left. +\frac{A_1^{0c}+A_0^{1c}-A_0^{0c}-A_1^{1c}}{8}
\left(c^1\downarrow u\uparrow u\uparrow
+u\uparrow u\uparrow c^1\downarrow\right)+\right.$$

$$\left. +\frac{A_1^{0c}-A_0^{1c}+A_0^{0c}-A_1^{1c}}{8}
\left(c^1\uparrow u\uparrow u\downarrow
+u\downarrow u\uparrow c^1\uparrow\right)
\right)\, ,\eqno (29)$$

$$k_{23}\left(\frac{A_1^{0c}}{2}\left(u^1\downarrow u\uparrow c\uparrow
+u^1\downarrow c\uparrow u\uparrow\right)
\right)\, .\eqno (30)$$

Then all members of wave function can be considered. After the grouping of
these members we can obtain:

$$u^1\downarrow u\uparrow c\uparrow \left\{
k_{12}\frac{A_1^0+3A_0^1}{4}+k_{13}\frac{A_1^{0c}+3A_0^{1c}}{4}
+k_{23}\, A_1^{0c}\right\}\, .\eqno (31)$$

The left side of the diagram (Fig.2) corresponds to the quark interactions.
The right side of Fig.2 determines the zero approximation (first diagram)
and the subsequent pair interactions (second diagram). The contribution to
$u^1\downarrow u\uparrow c\uparrow$ is shown in the Fig.3.

If we group the same members we obtain the system integral equations
for the $\Sigma_c$ state with the $J^p=\frac{3}{2} ^{-}$ $(10,2)$
multiplet:

$$\left\{
\begin{array}{l}
A_1^0(s,s_{12})=\lambda\, b_{1^+}(s_{12})L_{1^+}(s_{12})+
K_{1^+}(s_{12})\left[\frac{1}{4}A_1^{0c}(s,s_{13})+
\frac{3}{4}A_0^{1c}(s,s_{13})+\right.\\
\\
\hskip10ex \left.
+\frac{1}{4}A_1^{0c}(s,s_{23})+\frac{3}{4}A_0^{1c}(s,s_{23})
\right]\\
\\
A_1^{0c}(s,s_{13})=\lambda\, b_{1_c^+}(s_{13})L_{1_c^+}(s_{13})+
K_{1_c^+}(s_{13})\left[\frac{1}{2}A_1^0(s,s_{12})-
\frac{1}{4}A_1^{0c}(s,s_{12})+\right.\\
\\
\hskip10.5ex \left.
+\frac{3}{4}A_0^{1c}(s,s_{12})+\frac{1}{2}A_1^0(s,s_{23})-
\frac{1}{4}A_1^{0c}(s,s_{23})+\frac{3}{4}A_0^{1c}(s,s_{23})
\right] \\
\\
A_0^{1c}(s,s_{23})=\lambda\, b_{1_c^-}(s_{23})L_{1_c^-}(s_{23})+
K_{1_c^-}(s_{23})\left[\frac{1}{2}A_1^0(s,s_{12})+
\frac{1}{4}A_1^{0c}(s,s_{12})+\right.\\
\\
\hskip10.5ex \left.
+\frac{1}{4}A_0^{1c}(s,s_{12})+\frac{1}{2}A_1^0(s,s_{13})+
\frac{1}{4}A_1^{0c}(s,s_{13})+\frac{1}{4}A_0^{1c}(s,s_{13})
\right] \, .\\
\end{array}
\right.\eqno (32)$$

Here function $L_J(s_{ik})$ has the form

$$L_J(s_{ik})=\frac{G_J(s_{ik})}{1-B_J(s_{ik})}.\eqno (33)$$

The integral operator $K_J (s_{ik})$ is:

$$K_J (s_{ik})=L_J(s_{ik})\, \int\limits_{(m_i+m_k)^2}^{\infty}\hskip2mm
\frac{ds'_{ik}}{\pi}\frac{\rho_J(s'_{ik})G_J(s'_{ik})}
{s'_{ik}-s_{ik}}\, \int\limits_{-1}^{1}\frac{dz}{2}\, .\eqno (34)$$

$$b_J(s_{ik})=\int\limits_{(m_i+m_k)^2}^{\infty}\,
\frac{ds'_{ik}}{\pi}\frac{\rho_J(s'_{ik})G_J(s'_{ik})}
{s'_{ik}-s_{ik}}\, .\eqno (35)$$

The function $b_J(s_{ik})$ is the truncated function of Chew-Mandelstam.
$z$ is the cosine of the angle between the relative momentum of particles
$i$ and $k$ in the intermediate state and the momentum of
particle $j$ in the final state, taken in the c.m. of the particles
$i$ and $k$. Let some current produces three quarks (first diagram Fig.1)
with the vertex constant $\lambda$. This constant do not affect to the
spectra mass of excited baryons.

By analogy with the $\Sigma_c$ $\frac{3}{2} ^{-}$ $(10,2)$ state we obtain
the rescattering amplitudes of the three various quarks for all $P$-wave
states of the $(70,1^-)$ multiplet which satisfy the system of
integral equations.

\vskip2ex
{\bf 4. The reduced equations of ${\bf (70,1^-)}$ multiplet.}
\vskip2ex

Let us extract two-particle singularities in $A_J(s,s_{ik})$:

$$A_J(s,s_{ik})=\frac{\alpha_J(s,s_{ik})b_J(s_{ik})G_J(s_{ik})}
{1-B_J(s_{ik})},\eqno (36)$$

\noindent
$\alpha_J(s,s_{ik})$ is the reduced amplitude. Accordingly all integral
equations can be rewritten using the reduced amplitudes. For instance, one
consider the first equation of system for the $\Sigma_c$ $J^p=\frac{3}{2}^-$
of the $(10,2)$ multiplet:

$$\alpha_1^0 (s,s_{12})=\lambda+\frac{1}{b_{1^+}(s_{12})}
\, \int\limits_{(m_1+m_2)^2}^{\Lambda_{1^+}(1,2)}\,
\frac{ds'_{12}}{\pi}\,\frac{\rho_{1^+}(s'_{12})G_{1^+}(s'_{12})}
{s'_{12}-s_{12}}\times$$

$$\times\int\limits_{-1}^{1}\frac{dz}{2}\,
\left(
\frac{G_{1_c^+}(s'_{13})b_{1_c^+}(s'_{13})}{1-B_{1_c^+}(s'_{13})}
\,\frac{1}{2}\,\alpha_1^{0c}(s,s'_{13})+
\frac{G_{1_c^-}(s'_{13})b_{1_c^-}(s'_{13})}{1-B_{1_c^-}(s'_{13})}
\,\frac{3}{2}\,\alpha_0^{1c}(s,s'_{13})
\right).\eqno (37)$$

The connection between $s'_{12}$ and $s'_{13}$ is [20]:

$$s'_{13}=m_1^2+m_3^2-\frac{\left(s'_{12}+m_3^2-s\right)
\left(s'_{12}+m_1^2-m_2^2\right)}{2s'_{12}}\pm$$

$$\pm\frac{z}{2s'_{12}}\times\sqrt{\left(s'_{12}-(m_1+m_2)^2\right)
\left(s'_{12}-(m_1-m_2)^2\right)}\times$$

$$\times\sqrt{\left(s'_{12}-(\sqrt{s}+m_3)^2\right)
\left(s'_{12}-(\sqrt{s}-m_3)^2\right)}\, .\eqno (38)$$

The formula for $s'_{23}$ is similar to (38) with $z$ replaced by $-z$.
Thus $A_1^{0c}(s,s'_{13})+A_1^{0c}(s,s'_{23})$ must be replaced by
$2A_1^{0c}(s,s'_{13})$. $\Lambda_J(i,k)$ is the cutoff at the large
value of $s_{ik}$, which determines the contribution from small distances.

The construction of the approximate solution of the (37) is based on
the extraction of the leading singularities which are close to the
region $s_{ik}=(m_i+m_k)^2$ [20]. Amplitudes with different number of
rescattering have the following structure of singularities. The main
singularities in $s_{ik}$ are from pair rescattering of the particles
$i$ and $k$. First of all there are threshold square root singularities.
Also possible are pole singularities, which correspond to the bound
states. The diagrams in Fig.2 apart from two-particle singularities
have their own specific triangle singularities. Such classification
allows us to search the approximate solution of (37) by taking into
account some definite number of leading singularities and neglecting all
the weaker ones.

We consider the approximation, which corresponds to the single interaction
of all three particles (two-particle and triangle singularities) and
neglecting all the weaker ones.

The functions $\alpha_J(s,s_{ik})$ are the smooth functions of $s_{ik}$
as compared with the singular part of the amplitude, hence it can be
expanded in a series in the singulary point and only the first term of
this series should be employed further. As $s_0$ it is convenient to
take the middle point of physical region of Dalitz-plot in which $z=0$.
In this case we get
$s_{ik}=s_0=\frac{s+m_1^2+m_2^2+m_3^2}{m_{12}^2+m_{13}^2+m_{23}^2}$,
where $m_{ik}=\frac{m_i+m_k}{2}$. We define $\alpha_J(s,s_{ik})$ and
$b_J(s_{ik})$ at the point $s_0$. Such a choice of point $s_0$ allows us
to replace integral equations (37) by the algebraic equations for the
state $\Sigma_c$ $J^p=\frac{3}{2}^-$ of the $(10,2)$ multiplet:

$$\left\{
\begin{array}{l}
\alpha_1^0(s,s_0)=\lambda+\frac{1}{2}\,\alpha_1^{0c}(s,s_0)
\, I_{1^+ 1^+_c}(s,s_0)\,\frac{b_{1^+_c}(s_0)}{b_{1^+}(s_0)}
+\frac{3}{2}\,\alpha_0^{1c}(s,s_0)\, I_{1^+ 1^-_c}(s,s_0)
\,\frac{b_{1^-_c}(s_0)}{b_{1^+}(s_0)}
\hskip5.5ex 1^+\\
\\
\alpha_1^{0c}(s,s_0)=\lambda+\alpha_1^0(s,s_0)\, I_{1^+_c 1^+}(s,s_0)
\,\frac{b_{1^+}(s_0)}{b_{1^+_c}(s_0)}-\frac{1}{2}\,\alpha_1^{0c}(s,s_0)
\, I_{1^+_c 1^+_c}(s,s_0)\hskip13.5ex 1^+_c\\
\\
\hskip10ex
+\frac{3}{2}\,\alpha_0^{1c}(s,s_0)\, I_{1^+_c 1^-_c}(s,s_0)
\,\frac{b_{1^-_c}(s_0)}{b_{1^+_c}(s_0)}\\
\\
\alpha_0^{1c}(s,s_0)=\lambda+\alpha_1^0(s,s_0)\, I_{1^-_c 1^+}(s,s_0)
\,\frac{b_{1^+}(s_0)}{b_{1^-_c}(s_0)}+\frac{1}{2}\,\alpha_1^{0c}(s,s_0)
\, I_{1^-_c 1^+_c}(s,s_0)\,\frac{b_{1^+_c}(s_0)}{b_{1^-_c}(s_0)}
\hskip7ex 1^-_c\\
\\
\hskip10ex
+\frac{1}{2}\,\alpha_0^{1c}(s,s_0)\, I_{1^-_c 1^-_c}(s,s_0)
\, .\\
\end{array} \right.\eqno (39)$$

Here the reduced amplitudes for the diquarks $1^+$, $1^+_c$, $1^-_c$
are given. The function $I_{J_1 J_2}(s,s_0)$ takes into account singularity
which corresponds to the simultaneous vanishing of all propagators in the
triangle diagrams.

$$I_{J_1 J_2}(s,s_0)=\int\limits_{(m_i+m_k)^2}^{\Lambda_{J_1}}\hskip2mm
\frac{ds'_{ik}}{\pi}\frac{\rho_{J_1}(s'_{ik})G^2_{J_1}(s'_{ik})}
{s'_{ik}-s_{ik}}\, \int\limits_{-1}^{1}\frac{dz}{2}\,
\frac{1}{1-B_{J_2}(s_{ij})}\, .\eqno (40)$$

The $G_J(s_{ik})$ functions have the smooth dependence from energy
$s_{ik}$ [18] therefore we suggest them as constants. The parameters of
model: $\lambda_J$ cutoff parameter, $g_J$ vertex constants are chosen
dimensionless:

$$g_J=\frac{m_i+m_k}{2\pi}G_J , \,\,\, \lambda_J=\frac{4\Lambda_J}
{(m_i+m_k)^2} .\eqno (41)$$

Here $m_i$ and $m_k$ are quark masses in the intermediate state of the quark
loop. We calculate the system equations and can determine the mass values
of the $\Sigma_c$ $J^p=\frac{3}{2}^-$ $(10,2)$. We calculate a pole in $s$
which corresponds to the bound state of three quarks.

By analogy with $\Sigma_c$-hyperon we obtain the system equations for the
reduced amplitudes for all particles $(70,1^-)$ multiplets.

\newpage
{\bf 5. Calculation results.}
\vskip2ex

The quark masses ($m_u=m_d=m$ and $m_c$) are not fixed. In any way we
assume $m=570\, MeV$ and $m_c=1900\, MeV$. The value of nonstrange mass
$m$ is similar to the our paper ones [15]. In our model the four parameters
are used: gluon coupling constants $g_c=0.85$ and $g_u=0.58$ and cutoff
energy parameters $\lambda_c=9.2$, $\lambda _u=10.2$
($\lambda_{cu}=\frac{1}{4}\left(\sqrt{\lambda_u}+\sqrt{\lambda_c}\right)^2$)
for the charmed and nonstrange diquarks. The parameters have been determined
by the baryon masses:
$M_{\Lambda_c \frac{1}{2}^- (8,4)}=2880\, MeV$,\,
$M_{\Lambda_c \frac{3}{2}^- (8,4)}=2625\, MeV$,\,
$M_{\Lambda_c \frac{5}{2}^- (8,4)}=2765\, MeV$, and
$M_{\Xi_{cc} \frac{5}{2}^- (8,4)}=3519\, MeV$.
In the Tables I-IV we represent the masses of the charmed resonances
belonging to the $(70,1^-)$ multiplet obtained using the fit of
experimental values [21].

The $(70,1^-)$ charmed baryon multiplet has $23$ baryons with different
masses. The $6$ resonances are in good agreement with the experimental data
[21]. We have predicted $17$ masses of charmed excited baryons.

In the framework of the proposed approximate method of solving the
relativistic three-particle problem, we have obtained a satisfactory
spectrum of $P$-wave charmed baryons.

\vskip2ex
{\bf 6. Conclusion.}
\vskip2ex

In strongly bound systems of light and heavy quarks, such as the charmed
baryons considered, where $p/m \sim 1$ for the light quarks, the
approximation by nonrelativistic kinematics and dynamics is not justified.
The relativized quark model applied to baryon spectroscopy by Capstick and
Isgur [22].

In the papers [13, 14] the relativistic generalization of Faddeev equations
in the framework of dispersion relations are constructed. We calculated
the $S$-wave baryon masses using the method based on the extraction of
leading singularities of the amplitude. The behavior of electromagnetic
form factor of the nucleon and hyperon in the region of low and
intermediate momentum transfers is determined by [23]. In the framework of
the dispersion relation approach the charge radii of $S$-wave baryon
multiplets with $J^p=\frac{1}{2}^+$ are calculated.

In our paper [24] the relativistic Faddeev equations for the $S$-wave
charmed baryons are constructed. We calculated the mass spectra of single,
double and triple charm baryons using the input four-fermion interaction
with quantum numbers of the gluon.

In the framework of a relativistically covariant constituent quark model
one calculated on the basic of the Bethe-Salpeter equation in its
instantaneous approximation mass spectrum of $P$-wave charmed baryons [25].

In our paper [15] the relativistic description of three particles
amplitudes of $P$-wave baryons are considered. We take into account
the $u, d, s$-quarks. The mass spectrum of nonstrange and strange states
of multiplet $(70,1^-)$ are calculated. We use only four parameters
for the calculation of $30$ baryon masses. We take into account the mass
shift of $u, d, s$ quarks which allows us to obtain the $P$-wave baryon
bound states [15]. Recently, the mass spectrum baryons of $(70,1^-)$
multiplet using $1/N_c$ expansion are calculated [26]. The authors solved
the problem by removing the splitting of generators and using
orbital-flavor-spin wave functions.

We also use the orbital-flavor-spin wave functions for the construction
of integral equations. It allows as to calculate the mass spectra for all
charmed baryons $(70,1^-)$ multiplet. The important problem is the mixing
of $P$-wave baryons and the five quark systems (cryptoexotic baryons) [27]
and hybrid baryons [28]. We can see that the masses of $P$-wave charmed
baryons with $J^p=\frac{1}{2}^-$ are heavier than the masses of states
with $J^p=\frac{3}{2}^-$ and $J^p=\frac{5}{2}^-$. This conclusion
contradicts to the result of nonrelativistic quark models [29 -- 32].
The exceptions are the masses of lowest $\Lambda_c$-baryons with
$J^p=\frac{1}{2}^-$ and $J^p=\frac{3}{2}^-$. The lowest state $\Lambda_c$
$(1,2)$ $J^p=\frac{1}{2}^-$ $(70,1^-)$ mass is equal to $M=2400\, MeV$.

\vskip2ex
{\bf Acknowledgment}
\vskip2ex

The authors would like to thank T. Barnes, S. Capstick, S. Chekanov,
Fl. Stancu for useful discussions. The work was carried with the support
of the Russion Ministry of Education (grant 2.1.1.68.26).

\vskip40pt

\noindent
{\large Table I.}

\noindent
{The $\Sigma_c$-hyperon masses of multiplet $(70,1^-)$.}

\vskip1.5ex

\noindent
\begin{tabular}{|c|c|c|c|}
\hline
Multiplet & Baryon & Mass ($GeV$) & Mass ($GeV$) (exp.) \\
\hline
$\frac{3}{2}^-$ $(10,2)$ & $D_{13}$ & $2.570$ & --\\
\hline
$\frac{1}{2}^-$ $(10,2)$ & $S_{11}$ & $2.915$ & --\\
\hline
$\frac{5}{2}^-$ $(8,4)$ & $D_{15}$ & $2.740$ & $2.800$\\
\hline
$\frac{3}{2}^-$ $(8,4)$ & $D_{13}$ & $2.570$ & --\\
\hline
$\frac{1}{2}^-$ $(8,4)$ & $S_{11}$ & $2.915$ & --\\
\hline
$\frac{3}{2}^-$ $(8,2)$ & $D_{13}$ & $2.575$ & --\\
\hline
$\frac{1}{2}^-$ $(8,2)$ & $S_{11}$ & $2.700$ & --\\
\hline
\end{tabular}

\vskip1.5ex

\noindent
{The parameters of model (Tables I-IV): gluon coupling constants
$g_c=0.85$, $g_u=0.58$,\\
cutoff energy parameters $\lambda_c=9.2$, $\lambda_u=10.4$.}

\vskip40pt

\noindent
{\large Table II.}

\noindent
{The $\Lambda_c$-hyperon masses of multiplet $(70,1^-)$.}

\vskip1.5ex

\noindent
\begin{tabular}{|c|c|c|c|}
\hline
Multiplet & Baryon & Mass ($GeV$) & Mass ($GeV$) (exp.) \\
\hline
$\frac{5}{2}^-$ $(8,4)$ & $D_{05}$ & $2.765$ & $2.765$ \\
\hline
$\frac{3}{2}^-$ $(8,4)$ & $D_{03}$ & $2.625$ &  $2.625$ \\
\hline
$\frac{1}{2}^-$ $(8,4)$ & $S_{01}$ & $2.880$ & $2.880$ \\
\hline
$\frac{3}{2}^-$ $(8,2)$ & $D_{03}$ & $2.630$ & --\\
\hline
$\frac{1}{2}^-$ $(8,2)$ & $S_{01}$ & $2.635$ & $2.595$ \\
\hline
$\frac{3}{2}^-$ $(1,2)$ & $D_{03}$ & $2.630$ & --\\
\hline
$\frac{1}{2}^-$ $(1,2)$ & $S_{01}$ & $2.400$ & --\\
\hline
\end{tabular}

\newpage

\noindent
{\large Table III.}

\noindent
{The $\Xi_{cc}$-hyperon masses of multiplet $(70,1^-)$.}

\vskip1.5ex

\noindent
\begin{tabular}{|c|c|c|c|}
\hline
Multiplet & Baryon & Mass ($GeV$) & Mass ($GeV$) (exp.) \\
\hline
$\frac{3}{2}^-$ $(10,2)$ & $D_{13}$ & $3.140$ & --\\
\hline
$\frac{1}{2}^-$ $(10,2)$ & $S_{11}$ & $3.850$ & --\\
\hline
$\frac{5}{2}^-$ $(8,4)$ & $D_{15}$ & $3.519$ & $3.519$ \\
\hline
$\frac{3}{2}^-$ $(8,4)$ & $D_{13}$ & $3.140$ & --\\
\hline
$\frac{1}{2}^-$ $(8,4)$ & $S_{11}$ & $3.850$ & --\\
\hline
$\frac{3}{2}^-$ $(8,2)$ & $D_{13}$ & $3.240$ & --\\
\hline
$\frac{1}{2}^-$ $(8,2)$ & $S_{11}$ & $3.410$ & --\\
\hline
\end{tabular}

\vskip40pt

\noindent
{\large Table IV.}

\noindent
{The $\Omega_{ccc}$-hyperon masses of multiplet $(70,1^-)$.}

\vskip1.5ex

\noindent
\begin{tabular}{|c|c|c|c|}
\hline
Multiplet & Baryon & Mass ($GeV$) & Mass ($GeV$) (exp.) \\
\hline
$\frac{3}{2}^-$ $(10,2)$ & $D_{03}$ & $3.470$ & --\\
\hline
$\frac{1}{2}^-$ $(10,2)$ & $S_{01}$ & $4.585$ & --\\
\hline
\end{tabular}

\vskip50pt

\noindent
{\large Table V. Coefficient of Ghew-Mandelstam functions for the
different diquarks.}

\vskip1.5ex

\begin{tabular}{|c|c|c|c|}
\hline
 &$\alpha_J$&$\beta_J$&$\delta_J$\\
\hline
 & & & \\
$1^+$&$\frac{1}{3}$&$\frac{4m_i m_k}{3(m_i+m_k)^2}-\frac{1}{6}$
&$-\frac{1}{6}(m_i-m_k)^2$\\
 & & & \\
$0^+$&$\frac{1}{2}$&$-\frac{1}{2}\frac{(m_i-m_k)^2}{(m_i+m_k)^2}$&0\\
 & & & \\
$0^-$&$0$&$\frac{1}{2}$&$-\frac{1}{2}(m_i-m_k)^2$\\
 & & & \\
$1^-$&$\frac{1}{2}$&$-\frac{1}{2}\frac{(m_i-m_k)^2}{(m_i+m_k)^2}$&0\\
 & & & \\
$2^-$&$\frac{3}{10}$&$\frac{1}{5}
\left(1-\frac{3}{2}\frac{(m_i-m_k)^2}{(m_i+m_k)^2}\right)$
&$-\frac{1}{5}(m_i-m_k)^2$\\
 & & & \\
\hline
\end{tabular}

\newpage

\vskip60pt
\begin{picture}(600,60)
\put(-10,40){\line(1,0){33}}
\put(-10,50){\line(1,0){28}}
\put(-10,60){\line(1,0){33}}
\put(19,46){\line(1,1){15}}
\put(22,41){\line(1,1){17}}
\put(27.5,38.5){\line(1,1){14}}
\put(41,56){\vector(2,1){28}}
\put(42.5,50){\vector(1,0){35}}
\put(41,44){\vector(2,-1){28}}
\put(30,50){\circle{25}}
\put(70,78){$1$}
\put(70,55){$2$}
\put(70,20){$3$}
\put(87,47){$=$}
\put(107,53){\line(1,0){28}}
\put(107,50){\line(1,0){28}}
\put(107,47){\line(1,0){28}}
\put(135,53){\vector(2,1){28}}
\put(135,50){\vector(1,0){35}}
\put(135,47){\vector(2,-1){28}}
\put(163,78){$1$}
\put(163,55){$2$}
\put(163,20){$3$}
\put(180,47){$+$}
\put(200,40){\line(1,0){33}}
\put(200,50){\line(1,0){28}}
\put(200,60){\line(1,0){33}}
\put(229,46){\line(1,1){15}}
\put(232,41){\line(1,1){17}}
\put(237.5,38.5){\line(1,1){14}}
\put(251,44){\vector(2,-1){28}}
\put(240,50){\circle{25}}
\put(268,54){\oval(33,33)[tl]}
\put(252,70){\oval(33,33)[br]}
\put(269,71){\vector(2,3){15}}
\put(269,71){\vector(2,-1){24}}
\put(287,95){$1$}
\put(295,65){$2$}
\put(280,20){$3$}
\put(300,47){$+$}
\end{picture}

\vskip60pt
\begin{picture}(600,60)
\put(90,47){$+$}
\put(110,40){\line(1,0){33}}
\put(110,50){\line(1,0){28}}
\put(110,60){\line(1,0){33}}
\put(139,46){\line(1,1){15}}
\put(142,41){\line(1,1){17}}
\put(147.5,38.5){\line(1,1){14}}
\put(161,44){\vector(2,-1){28}}
\put(150,50){\circle{25}}
\put(178,54){\oval(33,33)[tl]}
\put(162,70){\oval(33,33)[br]}
\put(179,71){\vector(2,3){15}}
\put(179,71){\vector(2,-1){24}}
\put(197,95){$1$}
\put(205,65){$3$}
\put(190,20){$2$}
\put(210,47){$+$}
\put(230,40){\line(1,0){33}}
\put(230,50){\line(1,0){28}}
\put(230,60){\line(1,0){33}}
\put(259,46){\line(1,1){15}}
\put(262,41){\line(1,1){17}}
\put(267.5,38.5){\line(1,1){14}}
\put(281,44){\vector(2,-1){28}}
\put(270,50){\circle{25}}
\put(298,54){\oval(33,33)[tl]}
\put(282,70){\oval(33,33)[br]}
\put(299,71){\vector(2,3){15}}
\put(299,71){\vector(2,-1){24}}
\put(317,95){$2$}
\put(325,65){$3$}
\put(310,20){$1$}
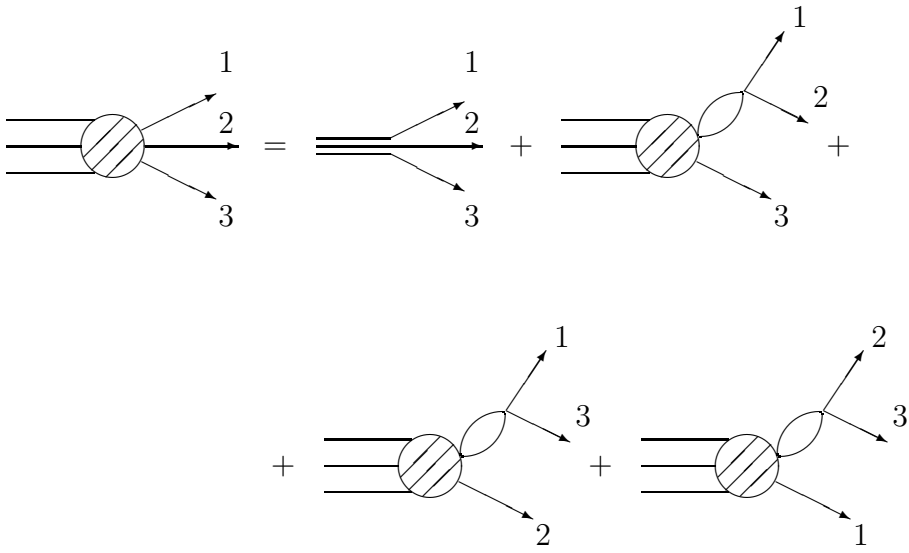
\put(-10,0){{\large Fig.1. The contribution of diagrams at the last pair
of the interacting particles.}}
\end{picture}

\vskip60pt

\vskip60pt
\begin{picture}(600,60)
\put(-10,40){\line(1,0){33}}
\put(-10,50){\line(1,0){28}}
\put(-10,60){\line(1,0){33}}
\put(19,46){\line(1,1){15}}
\put(22,41){\line(1,1){17}}
\put(27.5,38.5){\line(1,1){14}}
\put(41,44){\vector(2,-1){28}}
\put(30,50){\circle{25}}
\put(58,54){\oval(33,33)[tl]}
\put(42,70){\oval(33,33)[br]}
\put(59,71){\vector(2,3){15}}
\put(59,71){\vector(2,-1){24}}
\put(77,95){$1$}
\put(85,65){$2$}
\put(70,20){$3$}
\put(90,47){$=$}
\put(110,52){\line(1,0){28}}
\put(110,50){\line(1,0){28}}
\put(110,48){\line(1,0){28}}
\put(154,51){\oval(33,33)[tl]}
\put(138,67){\oval(33,33)[br]}
\put(155,67){\vector(2,3){15}}
\put(155,67){\vector(2,-1){24}}
\put(139,49){\vector(2,-1){28}}
\put(173,91){$1$}
\put(181,61){$2$}
\put(168,25){$3$}
\put(190,47){$+$}
\put(210,40){\line(1,0){33}}
\put(210,50){\line(1,0){28}}
\put(210,60){\line(1,0){33}}
\put(239,46){\line(1,1){15}}
\put(242,41){\line(1,1){17}}
\put(247.5,38.5){\line(1,1){14}}
\put(261,44){\vector(1,0){43}}
\put(250,50){\circle{25}}
\put(278,54){\oval(33,33)[tl]}
\put(262,70){\oval(33,33)[br]}
\put(279,71){\vector(2,3){15}}
\put(279,71){\vector(1,-1){25}}
\put(297,95){$3$}
\put(300,60){$1$}
\put(290,27){$2$}
\put(305,29){\oval(33,33)[tr]}
\put(321,45){\oval(33,33)[bl]}
\put(323,29){\vector(2,3){15}}
\put(323,29){\vector(2,-1){24}}
\put(341,53){$1$}
\put(333,7){$2$}
\put(360,47){$+$}
\end{picture}

\vskip60pt
\begin{picture}(600,60)
\put(190,47){$+$}
\put(210,40){\line(1,0){33}}
\put(210,50){\line(1,0){28}}
\put(210,60){\line(1,0){33}}
\put(239,46){\line(1,1){15}}
\put(242,41){\line(1,1){17}}
\put(247.5,38.5){\line(1,1){14}}
\put(261,44){\vector(1,0){43}}
\put(250,50){\circle{25}}
\put(278,54){\oval(33,33)[tl]}
\put(262,70){\oval(33,33)[br]}
\put(279,71){\vector(2,3){15}}
\put(279,71){\vector(1,-1){25}}
\put(297,95){$3$}
\put(300,60){$2$}
\put(290,27){$1$}
\put(305,29){\oval(33,33)[tr]}
\put(321,45){\oval(33,33)[bl]}
\put(323,29){\vector(2,3){15}}
\put(323,29){\vector(2,-1){24}}
\put(341,53){$2$}
\put(333,7){$1$}
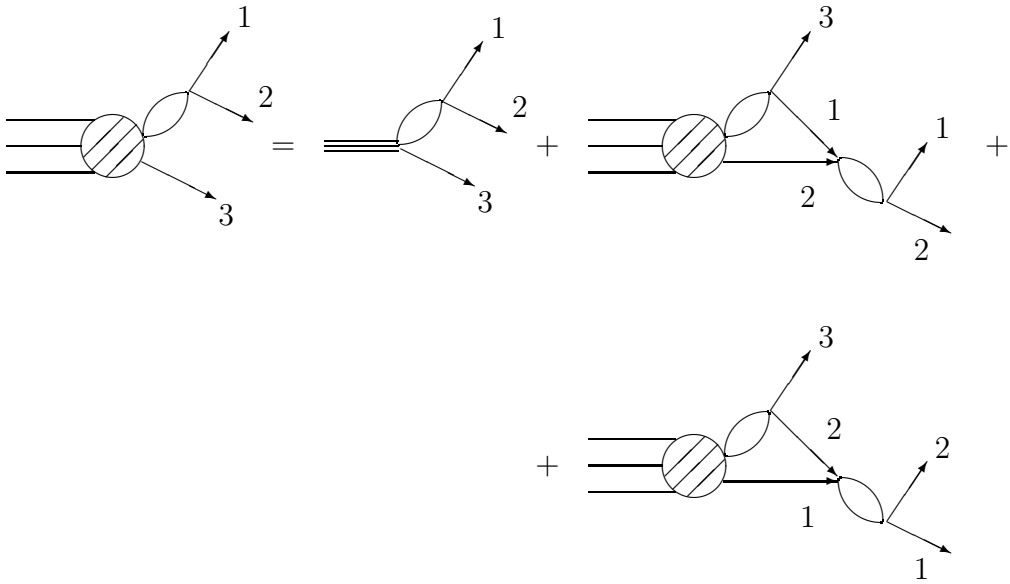
\put(-10,-10){{\large Fig.2. Graphic representation of the equations
for the amplitude $A_1(s,s_{ik})$.}}
\end{picture}

\newpage
\vskip60pt
\begin{picture}(600,80)
\multiput(70,20)(0,5){20}{\line(0,1){2}}
\put(-10,50){\line(1,0){33}}
\put(-10,60){\line(1,0){28}}
\put(-10,70){\line(1,0){33}}
\put(19,56){\line(1,1){15}}
\put(22,51){\line(1,1){17}}
\put(27.5,48.5){\line(1,1){14}}
\put(41,54){\vector(1,0){43}}
\put(30,60){\circle{25}}
\put(58,64){\oval(33,33)[tl]}
\put(42,80){\oval(33,33)[br]}
\put(59,81){\vector(2,3){15}}
\put(59,81){\vector(1,-1){25}}
\put(77,105){$3$}
\put(75,70){$1, 2$}
\put(60,37){$2, 1$}
\put(90,90){$k_{13}, k_{23}$}
\put(85,39){\oval(33,33)[tr]}
\put(101,55){\oval(33,33)[bl]}
\put(103,39){\vector(2,3){15}}
\put(103,39){\vector(2,-1){24}}
\put(121,40){$k_{12}\frac{A_1^0+3A_0^1}{4}\left|_{k_{12}}\right.$}
\put(200,70){$+$}
\multiput(305,20)(0,5){20}{\line(0,1){2}}
\put(225,50){\line(1,0){33}}
\put(225,60){\line(1,0){28}}
\put(225,70){\line(1,0){33}}
\put(254,56){\line(1,1){15}}
\put(257,51){\line(1,1){17}}
\put(262.5,48.5){\line(1,1){14}}
\put(276,54){\vector(1,0){43}}
\put(265,60){\circle{25}}
\put(293,64){\oval(33,33)[tl]}
\put(277,80){\oval(33,33)[br]}
\put(294,81){\vector(2,3){15}}
\put(294,81){\vector(1,-1){25}}
\put(312,105){$2$}
\put(310,70){$1, 3$}
\put(295,37){$3, 1$}
\put(325,90){$k_{12}, k_{23}$}
\put(320,39){\oval(33,33)[tr]}
\put(336,55){\oval(33,33)[bl]}
\put(338,39){\vector(2,3){15}}
\put(338,39){\vector(2,-1){24}}
\put(356,40){$k_{13}\frac{A_1^{0c}+3A_0^{1c}}{4}\left|_{k_{13}}\right.$}
\put(430,70){$+$}
\end{picture}

\vskip60pt
\begin{picture}(600,80)
\multiput(170,20)(0,5){20}{\line(0,1){2}}
\put(65,70){$+$}
\put(90,50){\line(1,0){33}}
\put(90,60){\line(1,0){28}}
\put(90,70){\line(1,0){33}}
\put(119,56){\line(1,1){15}}
\put(122,51){\line(1,1){17}}
\put(127.5,48.5){\line(1,1){14}}
\put(141,54){\vector(1,0){43}}
\put(130,60){\circle{25}}
\put(158,64){\oval(33,33)[tl]}
\put(142,80){\oval(33,33)[br]}
\put(159,81){\vector(2,3){15}}
\put(159,81){\vector(1,-1){25}}
\put(177,105){$1$}
\put(175,70){$2, 3$}
\put(160,37){$3, 2$}
\put(190,90){$k_{12}, k_{13}$}
\put(185,39){\oval(33,33)[tr]}
\put(201,55){\oval(33,33)[bl]}
\put(203,39){\vector(2,3){15}}
\put(203,39){\vector(2,-1){24}}
\put(221,40){$k_{23}A_1^{0c}\left|_{k_{23}}\right.$}
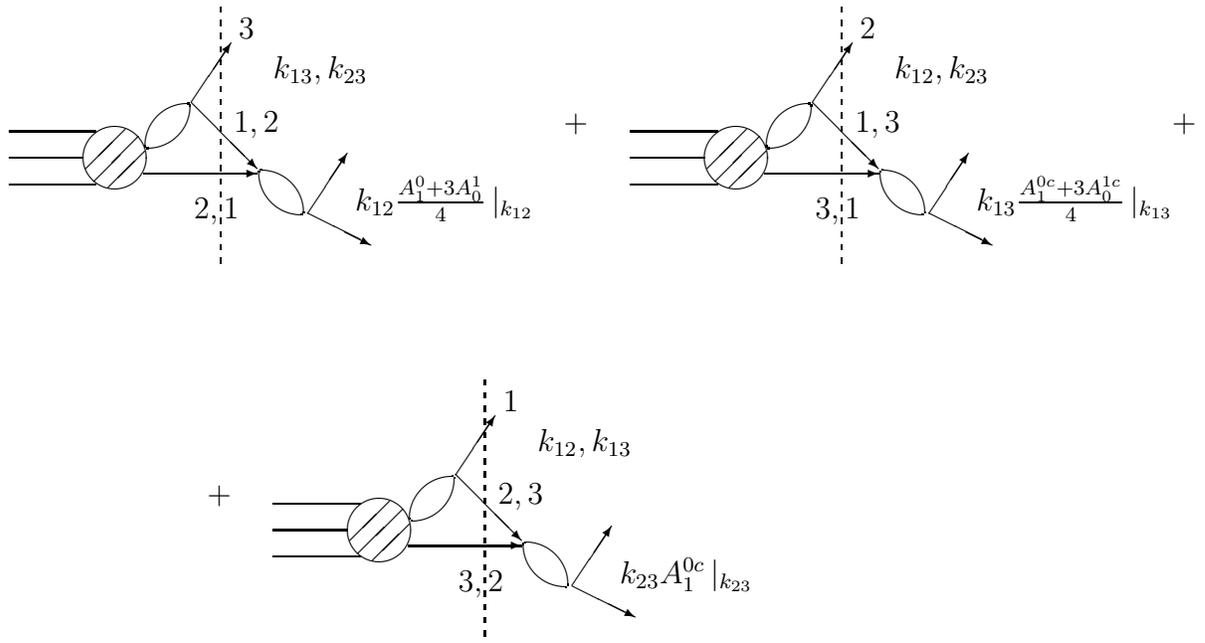
\put(-10,0){{\large Fig.3. The contribution of the diagrams with the
rescattering.}}
\end{picture}

\newpage
{\LARGE
{\bf
References.}}
\vskip5ex
\noindent
1. M. Artuso et el. (CLEO Collaboration), Phys. Rev. Lett. {\bf 86},
4479 (2001).

\noindent
2. R. Mizuk et el. (Belle Collaboration), Phys. Rev. Lett. {\bf 94},
122002 (2005).

\noindent
3. B. Aubert et el. (BABAR Collaboration), SLAC report SLAC-PUB-11786,

hep-ex/0603052.

\noindent
4. R. Chistov et el. (Belle Collaboration), Phys. Rev. Lett. {\bf 97},
162001 (2006).

\noindent
5. B. Aubert et el. (BABAR Collaboration), SLAC-PUB-11980,

BABAR-CONF-06-01, hep-ex/0607042.

\noindent
6. J.L. Rosner, hep-ph/0606166.

\noindent
7. J.L. Rosner, hep-ph/0612332.

\noindent
8. I.J.R. Aitchison, J. Phys. G{\bf 3}, 121 (1977).

\noindent
9. J.J. Brehm, Ann. Phys. (N.Y.) {\bf 108}, 454 (1977).

\noindent
10. I.J.R. Aitchison and J.J. Brehm, Phys. Rev. D{\bf 17}, 3072 (1978).

\noindent
11. I.J.R. Aitchison and J.J. Brehm,
Phys. Rev. D{\bf 20}, 1119, 1131 (1979).

\noindent
12. J.J. Brehm, Phys. Rev. D{\bf 21}, 718 (1980).

\noindent
13. S.M. Gerasyuta, Yad. Fiz. {\bf 55}, 3030 (1992).

\noindent
14. S.M. Gerasyuta, Z. Phys. C{\bf 60}, 683 (1993).

\noindent
15. S.M. Gerasyuta and E.E. Matskevich, Yad. Fiz. {\bf 70} (2007)
(in press),

hep-ph/0701120.

\noindent
16. F.E. Close, {\it An introduction to quarks and partons}, Academic
Press London

New York San Francisco, (1979) P. 438.

\noindent
17. A.De Rujula, H.Georgi and S.L.Glashow, Phys. Rev. D{\bf 12}, 147 (1975).

\noindent
18. V.V. Anisovich, S.M. Gerasyuta and A.V. Sarantsev,
Int. J. Mod. Phis. A{\bf 6}, 625

(1991).

\noindent
19. G. Chew and S. Mandelstam, Phys. Rev. {\bf 119}, 467 (1960).

\noindent
20. V.V. Anisovich and A.A. Anselm, Usp. Fiz. Nauk {\bf 88}, 287 (1966).

\noindent
21. W.M. Yao et al. (Particle Data Group), J. Phys. G{\bf 33}, 1 (2006).

\noindent
22. S. Capstick and N. Isgur, Phys. Rev. D{\bf 34}, 2809 (1986).

\noindent
23. S.M. Gerasyuta and D.V. Ivanov, Vest. SPb University Ser. 4, {\bf 11}, 3
(1996).

\noindent
24. S.M. Gerasyuta and D.V. Ivanov, Nuovo Cim. A{\bf 112}, 261 (1999).

\noindent
25. S. Migura, D. Merten, B. Metsch and H.-R. Petry, hep-ph/0602153.

\noindent
26. N. Matagne and Fl. Stancu, hep-ph/0610099.

\noindent
27. S.M. Gerasyuta and I.V. Kochkin,
Int. J. Mod. Phys. E{\bf 15}, 71 (2006).

\noindent
28. S.M. Gerasyuta and I.V. Kochkin, Phys. Rev. D{\bf 66}, 116001 (2002).

\noindent
29. K. Maltman and N. Isgur, Phys. Rev. D{\bf 22}, 1701 (1980).

\noindent
30. J.M. Richard and P. Taxil, Phys. Lett. B{\bf 128}, 453 (1983).

\noindent
31. C. Itoh, T. Minamikawa, K. Miura, T. Watahabe, Phys. Rev. D{\bf 40},
3660 (1989).

\noindent
32. S. Capstick and W. Roberts, Prog. Part. Nucl. Phys.
{\bf 45} S241 (2000).

\newpage
{\bf Appendix A. The ${\bf P}$-wave baryon wave functions.}
\vskip2ex
{\bf The wave functions of ${\bf (10,2)}$ decuplet.}
\vskip2ex

We considered this decuplet in the Section 2. The totally symmetric
$SU(6)\times O(3)$ wave function for each decuplet particle has the
following form:

$$\varphi=\frac{1}{\sqrt{2}}\left(
\varphi_{MA}^{SU(6)}\varphi_{MA}^{O(3)}+
\varphi_{MS}^{SU(6)}\varphi_{MS}^{O(3)}
\right)=\frac{1}{\sqrt{2}}\varphi_{S}^{SU(3)}\left(
\varphi_{MA}^{SU(2)}\varphi_{MA}^{O(3)}+
\varphi_{MS}^{SU(2)}\varphi_{MS}^{O(3)}
\right).\eqno (A1)$$

The functions $\varphi_{MA}^{SU(2)}$, $\varphi_{MS}^{SU(2)}$,
$\varphi_{MA}^{O(3)}$, $\varphi_{MS}^{O(3)}$ are:

$$\varphi_{MA}^{SU(2)}=\frac{1}{\sqrt{2}}\left(
\uparrow \downarrow \uparrow-\downarrow \uparrow \uparrow
\right),\quad
\varphi_{MS}^{SU(2)}=\frac{1}{\sqrt{6}}\left(
\uparrow \downarrow \uparrow+\downarrow \uparrow \uparrow-
2\uparrow \uparrow \downarrow
\right),\eqno (A2)$$

$$\varphi_{MA}^{O(3)}=\frac{1}{\sqrt{2}}\left(010-100\right),\quad
\varphi_{MS}^{O(3)}=\frac{1}{\sqrt{6}}\left(
010+100-2\cdot 001\right).\eqno (A3)$$

For the $\Sigma^{+}_c$-hyperon $SU(3)$-function is:

$$\varphi_{S}^{SU(3)}=\frac{1}{\sqrt{3}}\left(
ucu+cuu+uuc
\right).\eqno (A4)$$

Then one obtain the $SU(6)\times O(3)$-function of $\Sigma_c$ the
$(10,2)$ multiplet:

$$\varphi_{\Sigma^{+}_c(10,2)}=\frac{\sqrt{6}}{18}\left(
2\{u^1\downarrow u\uparrow c\uparrow\}+
\{c^1\downarrow u\uparrow u\uparrow\}-\right.$$
$$\left.-\{u^1\uparrow u\downarrow c\uparrow\}-
\{u^1\uparrow u\uparrow c\downarrow \}-
\{c^1\uparrow u\uparrow u\downarrow \}\right).\eqno (A5)$$

The replacement by $u\leftrightarrow c$ or $d\leftrightarrow c$ allows
us to obtain the $\Xi_{cc}$ wave function using the $\Sigma_c$ wave
function. In the case of $\Omega^{-}_{ccc}$ the $SU(6)\times O(3)$ wave
functions are given:

$$\varphi_{\Omega^{-}_{ccc}(10,2)}=\frac{\sqrt{2}}{6}\left(
\{c^1\downarrow c\uparrow c\uparrow\}
-\{c^1\uparrow c\uparrow c\downarrow\}
\right).\eqno (A6)$$

\vskip2ex
{\bf The wave functions of ${\bf (8,2)}$ octet.}
\vskip2ex
The wave functions of octet $\frac{3}{2} ^{-} ,\frac{1}{2} ^{-}$
$(8,2)$ multiplet are constructed as:

$$\varphi=\frac{1}{\sqrt{2}}\left(
\varphi_{MA}^{SU(6)}\varphi_{MA}^{O(3)}+
\varphi_{MS}^{SU(6)}\varphi_{MS}^{O(3)}
\right),\eqno (A7)$$

\noindent
here

$$\varphi_{MA}^{SU(6)}=\frac{1}{\sqrt{2}}\left(
\varphi_{MS}^{SU(3)}\varphi_{MA}^{SU(2)}+
\varphi_{MA}^{SU(3)}\varphi_{MS}^{SU(2)}
\right),\eqno (A8)$$

$$\varphi_{MS}^{SU(6)}=\frac{1}{\sqrt{2}}\left(
-\varphi_{MS}^{SU(3)}\varphi_{MS}^{SU(2)}+
\varphi_{MA}^{SU(3)}\varphi_{MA}^{SU(2)}
\right).\eqno (A9)$$

In the case of $\Sigma^{+}_c$ the $SU(3)$ wave functions
$\varphi_{MS}^{SU(3)}$ and $\varphi_{MA}^{SU(3)}$ have the following form:

$$\varphi_{MS}^{SU(3)}=\frac{1}{\sqrt{6}}\left(
ucu+cuu-2uuc\right),\quad
\varphi_{MA}^{SU(3)}=\frac{1}{\sqrt{2}}\left(
ucu-cuu\right).\eqno (A10)$$

Then we can obtain the symmetric wave function for $\Sigma^{+}_c$:

$$\varphi_{\Sigma^{+}_c (8,2)}=\frac{\sqrt{6}}{18}\left(
2\{u^1\uparrow u\downarrow c\uparrow\}+
\{c^1\downarrow u\uparrow u\uparrow\}-
\{u^1\uparrow u\uparrow c\downarrow\}-\right.$$

$$\left.-\{u^1\downarrow u\uparrow c\uparrow\}-
\{c^1\uparrow u\uparrow u\downarrow\}
\right).\eqno (A11)$$

The $\Xi^{0}_{cc}$ wave functions are obtained with replacement by
$u\leftrightarrow c$ in $\Sigma^{+}_c$.

The $\Lambda^{0}_c$ $SU(3)$ wave functions $\varphi_{MS}^{SU(3)}$ and
$\varphi_{MA}^{SU(3)}$ are given:

$$\varphi_{MS}^{SU(3)}=\frac{1}{2}\left(
dcu-ucd+cdu-cud
\right),\eqno (A12)$$

$$\varphi_{MA}^{SU(3)}=\frac{\sqrt{3}}{6}\left(
cdu-cud+ucd-dcu-2duc+2udc
\right).\eqno (A13)$$

Then the symmetric $SU(6)\times O(3)$ wave function for $\Lambda^{0}_c$
$\frac{3}{2} ^{-} ,\frac{1}{2} ^{-}$ can be considered as:

$$\varphi_{\Lambda^{0}_c (8,2)}=\frac{1}{6}\left(
\{u^1\uparrow d\uparrow c\downarrow\}-
\{u^1\downarrow d\uparrow c\uparrow\}-
\{d^1\uparrow u\uparrow c\downarrow\}+
\right.$$
$$\left.
+\{d^1\downarrow u\uparrow c\uparrow\}-
\{c^1\uparrow u\uparrow d\downarrow\}+
\{c^1\uparrow u\downarrow d\uparrow\}
\right).\eqno (A14)$$

\vskip2ex
{\bf The wave function of ${\bf (8,4)}$ octet.}
\vskip2ex
By analogy with the cases $(10,2)$ and $(8,2)$ we can calculate the
$(8,4)$ octet wave functions:

$$\varphi=\frac{1}{\sqrt{2}}\left(
\varphi_{MA}^{SU(6)}\varphi_{MA}^{O(3)}+
\varphi_{MS}^{SU(6)}\varphi_{MS}^{O(3)}
\right),\eqno (A15)$$

\noindent
here

$$\varphi_{MA}^{SU(6)}=
\varphi_{MA}^{SU(3)}\varphi_{S}^{SU(2)},\quad
\varphi_{MS}^{SU(6)}=
\varphi_{MS}^{SU(3)}\varphi_{S}^{SU(2)}.\eqno (A16)$$

$SU(2)$ wave function is totally symmetric:

$$\varphi_{S}^{SU(2)}=\uparrow\uparrow\uparrow ,\eqno (A17)$$

\noindent
$\varphi_{MS}^{SU(3)}$ and $\varphi_{MA}^{SU(3)}$ similar to one of the
$(8,2)$ multiplet.

For the $\Sigma^{+}_c$ $\frac{3}{2} ^{-} ,\frac{1}{2} ^{-}$ of $(8,4)$
multiplet one have:

$$\varphi_{\Sigma^{+}_c (8,4)}=\frac{\sqrt{2}}{6}\left(
\{c^1\uparrow u\uparrow u\uparrow\}-
\{u^1\uparrow u\uparrow c\uparrow\}
\right).\eqno (A18)$$

For the $\Xi^{0}_{cc}$ we can replace by $u\leftrightarrow c$ in
$\Sigma^{+}_c$.

We obtain the wave function of the $\Lambda^{0}_c$ $(8,4)$:

$$\varphi_{\Lambda^{0}_c (8,4)}=\frac{\sqrt{3}}{6}\left(
-\{u^1\uparrow d\uparrow c\uparrow\}+
\{d^1\uparrow u\uparrow c\uparrow\}
\right).\eqno (A19)$$

\vskip2ex
{\bf The ${\bf (1,2)}$ singlet wave function.}
\vskip2ex
We can use the totally symmetric $SU(6)\times O(3)$ wave function in the
form:

$$\varphi=\varphi_{A}^{SU(3)}\varphi_{A}^{SU(2)\times O(3)},\eqno (A20)$$

\noindent
here

$$\varphi_{A}^{SU(3)}=\frac{1}{\sqrt{6}}\left(
cdu-cud+ucd-dcu+duc-udc
\right),\eqno (A21)$$

$$\varphi_{A}^{SU(2)\times O(3)}=\frac{1}{\sqrt{2}}\left(
\varphi_{MS}^{SU(2)}\varphi_{MA}^{O(3)}-
\varphi_{MA}^{SU(2)}\varphi_{MS}^{O(3)}
\right).\eqno (A22)$$

As result we obtain:

$$\varphi_{\Lambda^{0}_c (1,2)}=\frac{\sqrt{3}}{6}\left(
-\{u^1\uparrow d\uparrow c\downarrow\}+
\{u^1\uparrow d\downarrow c\uparrow\}+
\{d^1\uparrow u\uparrow c\downarrow\}-\right.$$
$$\left.
-\{d^1\uparrow u\downarrow c\uparrow\}-
\{c^1\uparrow u\uparrow d\downarrow\}+
\{c^1\uparrow u\downarrow d\uparrow\}
\right).\eqno (A23)$$

\vskip2ex
{\bf Appendix B. The integral equations for the ${\bf (70,1^-)}$ multiplet.}
\vskip2ex
{\bf The ${\bf (10,2)}$ multiplet.}
\vskip2ex

We can represent the equations for the $\frac{3}{2}^-$ $(10,2)$ multiplet,
which is determined by the projection of orbital angular momentum $l_z=+1$.
We consider the following states:

$\Sigma_c$ $\frac{3}{2} ^{-}$:

$$\left\{
\begin{array}{l}
A_1^0(s,s_{12})=\lambda\, b_{1^+}(s_{12})L_{1^+}(s_{12})+
K_{1^+}(s_{12})\left[\frac{1}{4}A_1^{0c}(s,s_{13})+
\frac{3}{4}A_1^{0c}(s,s_{13})+\right.\\
\hskip10ex \left.
+\frac{1}{4}A_1^{0c}(s,s_{23})+\frac{3}{4}A_0^{1c}(s,s_{23})
\right]\\
A_1^{0c}(s,s_{13})=\lambda\, b_{1_c^+}(s_{13})L_{1_c^+}(s_{13})+
K_{1_c^+}(s_{13})\left[\frac{1}{2}A_1^0(s,s_{12})-
\frac{1}{4}A_1^{0c}(s,s_{12})+\right.\\
\hskip11ex \left.
+\frac{3}{4}A_0^{1c}(s,s_{12})+\frac{1}{2}A_1^0(s,s_{23})-
\frac{1}{4}A_1^{0c}(s,s_{23})+\frac{3}{4}A_0^{1c}(s,s_{23})
\right] \\
A_0^{1c}(s,s_{23})=\lambda\, b_{1_c^-}(s_{23})L_{1_c^-}(s_{23})+
K_{1_c^-}(s_{23})\left[\frac{1}{2}A_1^0(s,s_{12})+
\frac{1}{4}A_1^{0c}(s,s_{12})+\right.\\
\hskip11ex \left.
+\frac{1}{4}A_0^{1c}(s,s_{12})+\frac{1}{2}A_1^0(s,s_{13})+
\frac{1}{4}A_1^{0c}(s,s_{13})+\frac{1}{4}A_0^{1c}(s,s_{13})
\right] \, .\\
\end{array}
\right.\eqno (A24)$$

$\Omega_{ccc}$ $\frac{3}{2} ^{-}$:

$$\left\{
\begin{array}{l}
A_1^{0cc}(s,s_{12})=\lambda\, b_{1_{cc}^+}(s_{12})L_{1_{cc}^+}(s_{12})+
K_{1_{cc}^+}(s_{12})\left[\frac{1}{4}A_1^{0cc}(s,s_{13})+
\frac{3}{4}A_0^{1cc}(s,s_{13})+\right.\\
\hskip10ex \left.
+\frac{1}{4}A_1^{0cc}(s,s_{23})+\frac{3}{4}A_0^{1cc}(s,s_{23})
\right]\\
A_0^{1cc}(s,s_{13})=\lambda\, b_{1_{cc}^-}(s_{13})L_{1_{cc}^-}(s_{13})+
K_{1_{cc}^-}(s_{13})\left[\frac{3}{4}A_1^{0cc}(s,s_{12})+
\frac{1}{4}A_0^{1cc}(s,s_{12})+
\right.\\
\hskip10ex \left.
+\frac{3}{4}A_1^{0cc}(s,s_{23})+\frac{1}{4}A_0^{1cc}(s,s_{23})
\right] \, .\\
\end{array}
\right.\eqno (A25)$$

The system integral equations for the $\Xi_{cc}$ $\frac{3}{2} ^{-}$ are
similar to $\Sigma_c$ $\frac{3}{2} ^{-}$ with the replacement by
$u\leftrightarrow c$.

\newpage
{\bf The ${\bf (8,2)}$ multiplet.}
\vskip2ex

We determine the equations of multiplet $\frac{3}{2}^-$ $(8,2)$.

$\Sigma_c$ $\frac{3}{2} ^{-}$:

$$\left\{
\begin{array}{l}
A_1^0(s,s_{12})=\lambda\, b_{1^+}(s_{12})L_{1^+}(s_{12})+
K_{1^+}(s_{12})\left[-\frac{1}{8}A_1^{0c}(s,s_{13})+
\frac{3}{8}A_0^{1c}(s,s_{13})+
\right.\\
\hskip10.5ex
+\frac{3}{8}A_0^{0c}(s,s_{13})+\frac{3}{8}A_1^{1c}(s,s_{13})
-\frac{1}{8}A_1^{0c}(s,s_{23})+\frac{3}{8}A_0^{1c}(s,s_{23})+\\
\hskip10.5ex \left.
+\frac{3}{8}A_0^{0c}(s,s_{23})+\frac{3}{8}A_1^{1c}(s,s_{23})
\right]\\
A_1^{0c}(s,s_{13})=\lambda\, b_{1_c^+}(s_{13})L_{1_c^+}(s_{13})+
K_{1_c^+}(s_{13})\left[\frac{1}{2}A_1^0(s,s_{12})
-\frac{5}{8}A_1^{0c}(s,s_{12})+
\right.\\
\hskip11ex
+\frac{3}{8}A_0^{1c}(s,s_{12})+\frac{3}{8}A_0^{0c}(s,s_{12})
+\frac{3}{8}A_1^{1c}(s,s_{12})+\frac{1}{2}A_1^0(s,s_{23})-\\
\hskip11ex \left.
-\frac{5}{8}A_1^{0c}(s,s_{23})+\frac{3}{8}A_0^{1c}(s,s_{23})+
\frac{3}{8}A_0^{0c}(s,s_{23})+\frac{3}{8}A_1^{1c}(s,s_{23})
\right]\\
A_0^{1c}(s,s_{23})=\lambda\, b_{1_c^-}(s_{23})L_{1_c^-}(s_{23})+
K_{1_c^-}(s_{23})\left[\frac{1}{2}A_1^0(s,s_{12})
-\frac{1}{8}A_1^{0c}(s,s_{12})-
\right.\\
\hskip11ex
-\frac{1}{8}A_0^{1c}(s,s_{12})+\frac{3}{8}A_0^{0c}(s,s_{12})
+\frac{3}{8}A_1^{1c}(s,s_{12})+\frac{1}{2}A_1^0(s,s_{13})-\\
\hskip11ex \left.
-\frac{1}{8}A_1^{0c}(s,s_{13})-\frac{1}{8}A_0^{1c}(s,s_{13})+
\frac{3}{8}A_0^{0c}(s,s_{13})+\frac{3}{8}A_1^{1c}(s,s_{13})
\right]\\
A_0^{0c}(s,s_{13})=\lambda\, b_{0_c^+}(s_{13})L_{0_c^+}(s_{13})+
K_{0_c^+}(s_{13})\left[\frac{1}{2}A_1^0(s,s_{12})
-\frac{1}{8}A_1^{0c}(s,s_{12})+
\right.\\
\hskip11ex
+\frac{3}{8}A_0^{1c}(s,s_{12})-\frac{1}{8}A_0^{0c}(s,s_{12})
+\frac{3}{8}A_1^{1c}(s,s_{12})+\frac{1}{2}A_1^0(s,s_{23})-\\
\hskip11ex \left.
-\frac{1}{8}A_1^{0c}(s,s_{23})+\frac{3}{8}A_0^{1c}(s,s_{23})-
\frac{1}{8}A_0^{0c}(s,s_{23})+\frac{3}{8}A_1^{1c}(s,s_{23})
\right]\\
A_1^{1c}(s,s_{23})=\lambda\, b_{2_c^-}(s_{23})L_{2_c^-}(s_{23})+
K_{2_c^-}(s_{23})\left[\frac{1}{2}A_1^0(s,s_{12})
-\frac{1}{8}A_1^{0c}(s,s_{12})+
\right.\\
\hskip11ex
+\frac{3}{8}A_0^{1c}(s,s_{12})+\frac{3}{8}A_0^{0c}(s,s_{12})
-\frac{1}{8}A_1^{1c}(s,s_{12})+\frac{1}{2}A_1^0(s,s_{13})-\\
\hskip11ex \left.
-\frac{1}{8}A_1^{0c}(s,s_{13})+\frac{3}{8}A_0^{1c}(s,s_{13})+
\frac{3}{8}A_0^{0c}(s,s_{13})-\frac{1}{8}A_1^{1c}(s,s_{13})
\right]\, .\\
\end{array}
\right.\eqno (A26)$$

$\Lambda_c$ $\frac{3}{2} ^{-}$:

$$\left\{
\begin{array}{l}
A_0^1(s,s_{12})=\lambda\, b_{1^-}(s_{12})L_{1^-}(s_{12})+
K_{1^-}(s_{12})\left[\frac{3}{8}A_1^{0c}(s,s_{13})-
\frac{1}{8}A_0^{1c}(s,s_{13})+
\right.\\
\hskip10.5ex
+\frac{3}{8}A_0^{0c}(s,s_{13})+\frac{3}{8}A_1^{1c}(s,s_{13})
+\frac{3}{8}A_1^{0c}(s,s_{23})-\frac{1}{8}A_0^{1c}(s,s_{23})+\\
\hskip10.5ex \left.
+\frac{3}{8}A_0^{0c}(s,s_{23})+\frac{3}{8}A_1^{1c}(s,s_{23})
\right]\\
A_1^{0c}(s,s_{13})=\lambda\, b_{1_c^+}(s_{13})L_{1_c^+}(s_{13})+
K_{1_c^+}(s_{13})\left[\frac{1}{2}A_0^1(s,s_{12})
-\frac{1}{8}A_1^{0c}(s,s_{12})-
\right.\\
\hskip11ex
-\frac{1}{8}A_0^{1c}(s,s_{12})+\frac{3}{8}A_0^{0c}(s,s_{12})
+\frac{3}{8}A_1^{1c}(s,s_{12})+\frac{1}{2}A_0^1(s,s_{23})-\\
\hskip11ex \left.
-\frac{1}{8}A_1^{0c}(s,s_{23})-\frac{1}{8}A_0^{1c}(s,s_{23})+
\frac{3}{8}A_0^{0c}(s,s_{23})+\frac{3}{8}A_1^{1c}(s,s_{23})
\right]\\
A_0^{1c}(s,s_{23})=\lambda\, b_{1_c^-}(s_{23})L_{1_c^-}(s_{23})+
K_{1_c^-}(s_{23})\left[\frac{1}{2}A_0^1(s,s_{12})
+\frac{3}{8}A_1^{0c}(s,s_{12})-
\right.\\
\hskip11ex
-\frac{5}{8}A_0^{1c}(s,s_{12})+\frac{3}{8}A_0^{0c}(s,s_{12})
+\frac{3}{8}A_1^{1c}(s,s_{12})+\frac{1}{2}A_0^1(s,s_{13})+\\
\hskip11ex \left.
+\frac{3}{8}A_1^{0c}(s,s_{13})-\frac{5}{8}A_0^{1c}(s,s_{13})+
\frac{3}{8}A_0^{0c}(s,s_{13})+\frac{3}{8}A_1^{1c}(s,s_{13})
\right]\\
A_0^{0c}(s,s_{13})=\lambda\, b_{0_c^+}(s_{13})L_{0_c^+}(s_{13})+
K_{0_c^+}(s_{13})\left[\frac{1}{2}A_0^1(s,s_{12})
+\frac{3}{8}A_1^{0c}(s,s_{12})-
\right.\\
\hskip11ex
-\frac{1}{8}A_0^{1c}(s,s_{12})-\frac{1}{8}A_0^{0c}(s,s_{12})
+\frac{3}{8}A_1^{1c}(s,s_{12})+\frac{1}{2}A_0^1(s,s_{23})+\\
\hskip11ex \left.
+\frac{3}{8}A_1^{0c}(s,s_{23})-\frac{1}{8}A_0^{1c}(s,s_{23})-
\frac{1}{8}A_0^{0c}(s,s_{23})+\frac{3}{8}A_1^{1c}(s,s_{23})
\right]\\
A_1^{1c}(s,s_{23})=\lambda\, b_{2_c^-}(s_{23})L_{2_c^-}(s_{23})+
K_{2_c^-}(s_{23})\left[\frac{1}{2}A_0^1(s,s_{12})
+\frac{3}{8}A_1^{0c}(s,s_{12})-
\right.\\
\hskip11ex
-\frac{1}{8}A_0^{1c}(s,s_{12})+\frac{3}{8}A_0^{0c}(s,s_{12})
-\frac{1}{8}A_1^{1c}(s,s_{12})+\frac{1}{2}A_0^1(s,s_{13})+\\
\hskip11ex \left.
+\frac{3}{8}A_1^{0c}(s,s_{13})-\frac{1}{8}A_0^{1c}(s,s_{13})+
\frac{3}{8}A_0^{0c}(s,s_{13})-\frac{1}{8}A_1^{1c}(s,s_{13})
\right]\, .\\
\end{array}
\right.\eqno (A27)$$

The system equations of the $\Xi_{cc}$ $\frac{3}{2} ^{-}$ $(8,2)$ are
similar to the case $\Sigma_c$ $\frac{3}{2} ^{-}$ $(8,2)$ by replacement
$u\leftrightarrow c$.

\newpage
{\bf The ${\bf (8,4)}$ multiplet.}
\vskip2ex

We consider the states:

$\Sigma_c$ $\frac{5}{2} ^{-}$:

$$\left\{
\begin{array}{l}
A_1^0(s,s_{12})=\lambda\, b_{1^+}(s_{12})L_{1^+}(s_{12})+
K_{1^+}(s_{12})\left[\frac{1}{4}A_1^{0c}(s,s_{13})+
\frac{3}{4}A_1^{1c}(s,s_{13})+\right.\\
\hskip10ex \left.
+\frac{1}{4}A_1^{0c}(s,s_{23})+\frac{3}{4}A_1^{1c}(s,s_{23})
\right]\\
A_1^{0c}(s,s_{13})=\lambda\, b_{1_c^+}(s_{13})L_{1_c^+}(s_{13})+
K_{1_c^+}(s_{13})\left[\frac{1}{2}A_1^0(s,s_{12})-
\frac{1}{4}A_1^{0c}(s,s_{12})+\right.\\
\hskip11ex \left.
+\frac{3}{4}A_1^{1c}(s,s_{12})+\frac{1}{2}A_1^0(s,s_{23})-
\frac{1}{4}A_1^{0c}(s,s_{23})+\frac{3}{4}A_1^{1c}(s,s_{23})
\right] \\
A_1^{1c}(s,s_{23})=\lambda\, b_{2_c^-}(s_{23})L_{2_c^-}(s_{23})+
K_{2_c^-}(s_{23})\left[\frac{1}{2}A_1^0(s,s_{12})+
\frac{1}{4}A_1^{0c}(s,s_{12})+\right.\\
\hskip11ex \left.
+\frac{1}{4}A_1^{1c}(s,s_{12})+\frac{1}{2}A_1^0(s,s_{13})+
\frac{1}{4}A_1^{0c}(s,s_{13})+\frac{1}{4}A_1^{1c}(s,s_{13})
\right] \, .\\
\end{array}
\right.\eqno (A28)$$

$\Lambda_{cc}$ $\frac{5}{2} ^{-}$:

$$\left\{
\begin{array}{l}
A_1^1(s,s_{12})=\lambda\, b_{2^-}(s_{12})L_{2^-}(s_{12})+
K_{2^-}(s_{12})\left[\frac{3}{4}A_1^{0c}(s,s_{13})+
\frac{1}{4}A_1^{1c}(s,s_{13})+\right.\\
\hskip10ex \left.
+\frac{3}{4}A_1^{0c}(s,s_{23})+\frac{1}{4}A_1^{1c}(s,s_{23})
\right]\\
A_1^{0c}(s,s_{13})=\lambda\, b_{1_c^+}(s_{13})L_{1_c^+}(s_{13})+
K_{1_c^+}(s_{13})\left[\frac{1}{2}A_1^1(s,s_{12})+
\frac{1}{3}A_1^{0c}(s,s_{12})+\right.\\
\hskip11ex \left.
+\frac{1}{6}A_1^{1c}(s,s_{12})+\frac{1}{2}A_1^1(s,s_{23})+
\frac{1}{3}A_1^{0c}(s,s_{23})+\frac{1}{6}A_1^{1c}(s,s_{23})
\right] \\
A_1^{1c}(s,s_{23})=\lambda\, b_{2_c^-}(s_{23})L_{2_c^-}(s_{23})+
K_{2_c^-}(s_{23})\left[\frac{1}{2}A_1^1(s,s_{12})+
\frac{1}{2}A_1^{0c}(s,s_{12})+\right.\\
\hskip11ex \left.
+\frac{1}{2}A_1^1(s,s_{13})+
\frac{1}{2}A_1^{0c}(s,s_{13})
\right] \, .\\
\end{array}
\right.\eqno (A29)$$

\vskip2ex
{\bf The ${\bf (1,2)}$ multiplet.}
\vskip2ex

$\Lambda$ $\frac{3}{2} ^{-}$:

$$\left\{
\begin{array}{l}
A_0^0(s,s_{12})=\lambda\, b_{0^+}(s_{12})L_{0^+}(s_{12})+
K_{0^+}(s_{12})\left[\frac{1}{4}A_0^{0c}(s,s_{13})+
\frac{3}{4}A_1^{1c}(s,s_{13})+\right.\\
\hskip10ex \left.
+\frac{1}{4}A_0^{0c}(s,s_{23})+\frac{3}{4}A_1^{1c}(s,s_{23})
\right]\\
A_0^{0c}(s,s_{13})=\lambda\, b_{0_c^+}(s_{13})L_{0_c^+}(s_{13})+
K_{0_c^+}(s_{13})\left[\frac{1}{2}A_0^0(s,s_{12})-
\frac{1}{4}A_0^{0c}(s,s_{12})+\right.\\
\hskip11ex \left.
+\frac{3}{4}A_1^{1c}(s,s_{12})+\frac{1}{2}A_0^0(s,s_{23})-
\frac{1}{4}A_0^{0c}(s,s_{23})+\frac{3}{4}A_1^{1c}(s,s_{23})
\right] \\
A_1^{1c}(s,s_{23})=\lambda\, b_{2_c^-}(s_{23})L_{2_c^-}(s_{23})+
K_{2_c^-}(s_{23})\left[\frac{1}{2}A_0^0(s,s_{12})+
\frac{1}{4}A_0^{0c}(s,s_{12})+\right.\\
\hskip11ex \left.
+\frac{1}{4}A_1^{1c}(s,s_{12})+\frac{1}{2}A_0^0(s,s_{13})+
\frac{1}{4}A_0^{0c}(s,s_{13})+\frac{1}{4}A_1^{1c}(s,s_{13})
\right] \, .\\
\end{array}
\right.\eqno (A30)$$

\newpage
{\bf Appendix C. The system equations of reduced amplitude of the
multiplets ${\bf (70,1^-)}$.}
\vskip2ex
{\bf The equations of ${\bf (10,2)}$ multiplet.}
\vskip2ex

$\Sigma_c$ $\frac{3}{2} ^{-}$:

$$\left\{
\begin{array}{l}
\alpha_1^0(s,s_0)=\lambda+\frac{1}{2}\,\alpha_1^{0c}(s,s_0)
\, I_{1^+ 1^+_c}(s,s_0)\,\frac{b_{1^+_c}(s_0)}{b_{1^+}(s_0)}
+\frac{3}{2}\,\alpha_0^{1c}(s,s_0)\, I_{1^+ 1^-_c}(s,s_0)
\,\frac{b_{1^-_c}(s_0)}{b_{1^+}(s_0)}\hskip6ex 1^+\\
\alpha_1^{0c}(s,s_0)=\lambda+\alpha_1^0(s,s_0)\, I_{1^+_c 1^+}(s,s_0)
\,\frac{b_{1^+}(s_0)}{b_{1^+_c}(s_0)}-\frac{1}{2}\,\alpha_1^{0c}(s,s_0)
\, I_{1^+_c 1^+_c}(s,s_0)
\hskip14ex 1^+_c\\
\hskip10ex
+\frac{3}{2}\,\alpha_0^{1c}(s,s_0)\, I_{1^+_c 1^-_c}(s,s_0)
\,\frac{b_{1^-_c}(s_0)}{b_{1^+_c}(s_0)}\\
\alpha_0^{1c}(s,s_0)=\lambda+\alpha_1^0(s,s_0)\, I_{1^-_c 1^+}(s,s_0)
\,\frac{b_{1^+}(s_0)}{b_{1^-_c}(s_0)}+\frac{1}{2}\,\alpha_1^{0c}(s,s_0)
\, I_{1^-_c 1^+_c}(s,s_0)\,\frac{b_{1^+_c}(s_0)}{b_{1^-_c}(s_0)}
\hskip7.5ex 1^-_c\\
\hskip10ex
+\frac{1}{2}\,\alpha_0^{1c}(s,s_0)\, I_{1^-_c 1^-_c}(s,s_0)\, .\\
\end{array} \right.\eqno (A31)$$

$\Omega_{ccc}$ $\frac{3}{2} ^{-}$:

$$\left\{
\begin{array}{l}
\alpha_1^{0cc}(s,s_0)=\lambda+\frac{1}{2}\,\alpha_1^{0cc}(s,s_0)
\, I_{1_{cc}^+ 1_{cc}^+}(s,s_0)+\frac{3}{2}\,
\alpha_0^{1cc}(s,s_0)\, I_{1_{cc}^+ 1_{cc}^-}(s,s_0)\,
\frac{b_{1_{cc}^-}(s_0)}{b_{1_{cc}^+}(s_0)}\hskip8ex 1_{cc}^+\\
\alpha_0^{1cc}(s,s_0)=\lambda+\frac{3}{2}\,\alpha_1^{0cc}(s,s_0)
\, I_{1_{cc}^- 1_{cc}^+}(s,s_0)
\,\frac{b_{1_{cc}^+}(s_0)}{b_{1_{cc}^-}(s_0)}+\frac{1}{2}\,
\alpha_0^{1cc}(s,s_0)\,
I_{1_{cc}^- 1_{cc}^-}(s,s_0)\, . \hskip7ex 1_{cc}^-\\
\end{array} \right.\eqno (A32)$$

The $\Xi_{cc}$ system equations is similar to the case $\Sigma_c$ with the
replacement by $u\leftrightarrow c$.

\newpage
{\bf The equations of ${\bf (8,2)}$ multiplet.}
\vskip2ex

$\Sigma_c$ $\frac{3}{2} ^{-}$:

$$\left\{
\begin{array}{l}
\alpha_1^0(s,s_0)=\lambda-\frac{1}{4}\,\alpha_1^{0c}(s,s_0)
\, I_{1^+ 1^+_c}(s,s_0)\,\frac{b_{1^+_c}(s_0)}{b_{1^+}(s_0)}
+\frac{3}{4}\,\alpha_0^{1c}(s,s_0)\, I_{1^+ 1^-_c}(s,s_0)
\,\frac{b_{1^-_c}(s_0)}{b_{1^+}(s_0)}
\hskip4ex 1^+\\
\hskip9ex
+\frac{3}{4}\,\alpha_0^{0c}(s,s_0)\, I_{1^+ 0^+_c}(s,s_0)
\,\frac{b_{0^+_c}(s_0)}{b_{1^+}(s_0)}
+\frac{3}{4}\,\alpha_1^{1c}(s,s_0)\, I_{1^+ 2^-_c}(s,s_0)
\,\frac{b_{2^-_c}(s_0)}{b_{1^+}(s_0)}\\
\alpha_1^{0c}(s,s_0)=\lambda+\alpha_1^0(s,s_0)
\, I_{1^+_c 1^+}(s,s_0)\,\frac{b_{1^+}(s_0)}{b_{1^+_c}(s_0)}
-\frac{5}{4}\,\alpha_1^{0c}(s,s_0)\, I_{1^+_c 1^+_c}(s,s_0)
\hskip12.5ex 1^+_c\\
\hskip10ex
+\frac{3}{4}\,\alpha_0^{1c}(s,s_0)\, I_{1^+_c 1^-_c}(s,s_0)
\,\frac{b_{1^-_c}(s_0)}{b_{1^+_c}(s_0)}
+\frac{3}{4}\,\alpha_0^{0c}(s,s_0)\, I_{1^+_c 0^+_c}(s,s_0)
\,\frac{b_{0^+_c}(s_0)}{b_{1^+_c}(s_0)}\\
\hskip10ex
+\frac{3}{4}\,\alpha_1^{1c}(s,s_0)\, I_{1^+_c 2^-_c}(s,s_0)
\,\frac{b_{2^-_c}(s_0)}{b_{1^+_c}(s_0)}\\
\alpha_0^{1c}(s,s_0)=\lambda+\alpha_1^0(s,s_0)
\,I_{1^-_c 1^+}(s,s_0)\,\frac{b_{1^+}(s_0)}{b_{1^-_c}(s_0)}
-\frac{1}{4}\,\alpha_1^{0c}(s,s_0)\, I_{1^-_c 1^+_c}(s,s_0)
\,\frac{b_{1^+_c}(s_0)}{b_{1^-_c}(s_0)}\hskip6ex 1^-_c\\
\hskip10ex
-\frac{1}{4}\,\alpha_0^{1c}(s,s_0)\, I_{1^-_c 1^-_c}(s,s_0)
+\frac{3}{4}\,\alpha_0^{0c}(s,s_0)\, I_{1^-_c 0^+_c}(s,s_0)
\,\frac{b_{0^+_c}(s_0)}{b_{1^-_c}(s_0)}\\
\hskip10ex
+\frac{3}{4}\,\alpha_1^{1c}(s,s_0)\, I_{1^-_c 2^-_c}(s,s_0)
\,\frac{b_{2^-_c}(s_0)}{b_{1^-_c}(s_0)}\\
\alpha_0^{0c}(s,s_0)=\lambda+\alpha_1^0(s,s_0)
\, I_{0^+_c 1^+}(s,s_0)\,\frac{b_{1^+}(s_0)}{b_{0^+_c}(s_0)}
-\frac{1}{4}\,\alpha_1^{0c}(s,s_0)\, I_{0^+_c 1^+_c}(s,s_0)
\,\frac{b_{1^+_c}(s_0)}{b_{0^+_c}(s_0)}\hskip6ex 0^+_c\\
\hskip10ex
+\frac{3}{4}\,\alpha_0^{1c}(s,s_0)\, I_{0^+_c 1^-_c}(s,s_0)
\,\frac{b_{1^-_c}(s_0)}{b_{0^+_c}(s_0)}
-\frac{1}{4}\,\alpha_0^{0c}(s,s_0)\, I_{0^+_c 0^+_c}(s,s_0)\\
\hskip10ex
+\frac{3}{4}\,\alpha_1^{1c}(s,s_0)\, I_{0^+_c 2^-_c}(s,s_0)
\,\frac{b_{2^-_c}(s_0)}{b_{0^+_c}(s_0)}\\
\alpha_1^{1c}(s,s_0)=\lambda+\alpha_1^0(s,s_0)
\, I_{2^-_c 1^+}(s,s_0)\,\frac{b_{1^+}(s_0)}{b_{2^-_c}(s_0)}
-\frac{1}{4}\,\alpha_1^{0c}(s,s_0)\, I_{2^-_c 1^+_c}(s,s_0)
\,\frac{b_{1^+_c}(s_0)}{b_{2^-_c}(s_0)}\hskip6ex 2^-_c\\
\hskip10ex
+\frac{3}{4}\,\alpha_0^{1c}(s,s_0)\, I_{2^-_c 1^-_c}(s,s_0)
\,\frac{b_{1^-_c}(s_0)}{b_{2^-_c}(s_0)}
+\frac{3}{4}\,\alpha_0^{0c}(s,s_0)\, I_{2^-_c 0^+_c}(s,s_0)\\
\hskip10ex
\,\frac{b_{0^+_c}(s_0)}{b_{2^-_c}(s_0)}
-\frac{1}{4}\,\alpha_1^{1c}(s,s_0)\, I_{2^-_c 2^-_c}(s,s_0)\, .\\
\end{array} \right.\eqno (A33)$$

\newpage
$\Lambda_c$ $\frac{3}{2} ^{-}$:

$$\left\{
\begin{array}{l}
\alpha_1^0(s,s_0)=\lambda+\frac{3}{4}\,\alpha_1^{0c}(s,s_0)
\, I_{1^+ 1^+_c}(s,s_0)\,\frac{b_{1^+_c}(s_0)}{b_{1^+}(s_0)}
-\frac{1}{4}\,\alpha_0^{1c}(s,s_0)\, I_{1^+ 1^-_c}(s,s_0)
\,\frac{b_{1^-_c}(s_0)}{b_{1^+}(s_0)}\hskip4ex 1^+\\
\hskip9ex
+\frac{3}{4}\,\alpha_0^{0c}(s,s_0)\, I_{1^+ 0^+_c}(s,s_0)
\,\frac{b_{0^+_c}(s_0)}{b_{1^+}(s_0)}
+\frac{3}{4}\,\alpha_1^{1c}(s,s_0)\,I_{1^+ 2^-_c}(s,s_0)
\,\frac{b_{2^-_c}(s_0)}{b_{1^+}(s_0)}\\
\alpha_1^{0c}(s,s_0)=\lambda+\alpha_1^0(s,s_0)
\, I_{1^+_c 1^+}(s,s_0)\,\frac{b_{1^+}(s_0)}{b_{1^+_c}(s_0)}
-\frac{1}{4}\,\alpha_1^{0c}(s,s_0)\, I_{1^+_c 1^+_c}(s,s_0)
\hskip12.5ex 1^+_c\\
\hskip10ex
-\frac{1}{4}\,\alpha_0^{1c}(s,s_0)\, I_{1^+_c 1^-_c}(s,s_0)
\,\frac{b_{1^-_c}(s_0)}{b_{1^+_c}(s_0)}
+\frac{3}{4}\,\alpha_0^{0c}(s,s_0)\, I_{1^+_c 0^+_c}(s,s_0)
\,\frac{b_{0^+_c}(s_0)}{b_{1^+_c}(s_0)}\\
\hskip10ex
+\frac{3}{4}\,\alpha_1^{1c}(s,s_0)\, I_{1^+_c 2^-_c}(s,s_0)
\,\frac{b_{2^-_c}(s_0)}{b_{1^+_c}(s_0)}\\
\alpha_0^{1c}(s,s_0)=\lambda+\alpha_1^0(s,s_0)
\, I_{1^-_c 1^+}(s,s_0)\,\frac{b_{1^+}(s_0)}{b_{1^-_c}(s_0)}
+\frac{3}{4}\,\alpha_1^{0c}(s,s_0)\, I_{1^-_c 1^+_c}(s,s_0)
\,\frac{b_{1^+_c}(s_0)}{b_{1^-_c}(s_0)}\hskip6ex 1^-_c\\
\hskip10ex
-\frac{5}{4}\,\alpha_0^{1c}(s,s_0)\, I_{1^-_c 1^-_c}(s,s_0)
+\frac{3}{4}\,\alpha_0^{0c}(s,s_0)\, I_{1^-_c 0^+_c}(s,s_0)
\,\frac{b_{0^+_c}(s_0)}{b_{1^-_c}(s_0)}\\
\hskip10ex
+\frac{3}{4}\,\alpha_1^{1c}(s,s_0)\, I_{1^-_c 2^-_c}(s,s_0)
\,\frac{b_{2^-_c}(s_0)}{b_{1^-_c}(s_0)}\\
\alpha_0^{0c}(s,s_0)=\lambda+\alpha_1^0(s,s_0)
\, I_{0^+_c 1^+}(s,s_0)\,\frac{b_{1^+}(s_0)}{b_{0^+_c}(s_0)}
+\frac{3}{4}\,\alpha_1^{0c}(s,s_0)\, I_{0^+_c 1^+_c}(s,s_0)
\,\frac{b_{1^+_c}(s_0)}{b_{0^+_c}(s_0)}\hskip6ex 0^+_c\\
\hskip10ex
-\frac{1}{4}\,\alpha_0^{1c}(s,s_0)\, I_{0^+_c 1^-_c}(s,s_0)
\,\frac{b_{1^-_c}(s_0)}{b_{0^+_c}(s_0)}
-\frac{1}{4}\,\alpha_0^{0c}(s,s_0)\, I_{0^+_c 0^+_c}(s,s_0)\\
\hskip10ex
+\frac{3}{4}\,\alpha_1^{1c}(s,s_0)\, I_{0^+_c 2^-_c}(s,s_0)
\,\frac{b_{2^-_c}(s_0)}{b_{0^+_c}(s_0)}\\
\alpha_1^{1c}(s,s_0)=\lambda+\alpha_1^0(s,s_0)
\, I_{2^-_c 1^+}(s,s_0)\,\frac{b_{1^+}(s_0)}{b_{2^-_c}(s_0)}
+\frac{3}{4}\,\alpha_1^{0c}(s,s_0)\, I_{2^-_c 1^+_c}(s,s_0)
\,\frac{b_{1^+_c}(s_0)}{b_{2^-_c}(s_0)}\hskip6ex 2^-_c\\
\hskip10ex
-\frac{1}{4}\,\alpha_0^{1c}(s,s_0)\, I_{2^-_c 1^-_c}(s,s_0)
\,\frac{b_{1^-_c}(s_0)}{b_{2^-_c}(s_0)}
+\frac{3}{4}\,\alpha_0^{0c}(s,s_0)\, I_{2^-_c 0^+_c}(s,s_0)
\,\frac{b_{0^+_c}(s_0)}{b_{2^-_c}(s_0)}\\
\hskip10ex
-\frac{1}{4}\,\alpha_1^{1c}(s,s_0)\, I_{2^-_c 2^-_c}(s,s_0)\, .\\
\end{array} \right.\eqno (A34)$$

By analogy with the case $(10,2)$, the equations for the $\Xi_{cc}$
$\frac{3}{2} ^{-}$ $(8,2)$ are similar to $\Sigma_c$ $\frac{3}{2} ^{-}$
with replacement by $u\leftrightarrow c$.

\vskip2ex
{\bf The equations of ${\bf (8,4)}$ multiplet.}
\vskip2ex

$\Sigma_c$ $\frac{5}{2} ^{-}$:

$$\left\{
\begin{array}{l}
\alpha_1^0(s,s_0)=\lambda+\frac{1}{2}\,\alpha_1^{0c}(s,s_0)
\, I_{1^+ 1^+_c}(s,s_0)\,\frac{b_{1^+_c}(s_0)}{b_{1^+}(s_0)}
+\frac{3}{2}\,\alpha_1^{1c}(s,s_0)\, I_{1^+ 2^-_c}(s,s_0)
\,\frac{b_{2^-_c}(s_0)}{b_{1^+}(s_0)}\hskip4.5ex 1^+\\
\alpha_1^{0c}(s,s_0)=\lambda+\alpha_1^0(s,s_0)\, I_{1^+_c 1^+}(s,s_0)
\,\frac{b_{1^+}(s_0)}{b_{1^+_c}(s_0)}
-\frac{1}{2}\,\alpha_1^{0c}(s,s_0)\, I_{1^+_c 1^+_c}(s,s_0)
\hskip12.5ex 1^+_c\\
\hskip10ex
+\frac{3}{2}\,\alpha_1^{1c}(s,s_0)\, I_{1^+_c 2^-_c}(s,s_0)
\,\frac{b_{2^-_c}(s_0)}{b_{1^+_c}(s_0)}\\
\alpha_1^{1c}(s,s_0)=\lambda+\alpha_1^0(s,s_0)\, I_{2^-_c 1^+}(s,s_0)
\,\frac{b_{1^+}(s_0)}{b_{2^-_c}(s_0)}
+\frac{1}{2}\,\alpha_1^{0c}(s,s_0)\, I_{2^-_c 1^+_c}(s,s_0)
\,\frac{b_{1^+_c}(s_0)}{b_{2^-_c}(s_0)}\hskip6ex 2^-_c\\
\hskip10ex
+\frac{1}{2}\,\alpha_1^{1c}(s,s_0)\, I_{2^-_c 2^-_c}(s,s_0)\, .\\
\end{array} \right.\eqno (A35)$$

\newpage
$\Lambda_c$ $\frac{5}{2} ^{-}$:

$$\left\{
\begin{array}{l}
\alpha_1^1(s,s_0)=\lambda+\frac{3}{2}\,\alpha_1^{0c}(s,s_0)
\, I_{2^- 1^+_c}(s,s_0)
\,\frac{b_{1^+_c}(s_0)}{b_{2^-}(s_0)}
+\frac{1}{2}\,\alpha_1^{1c}(s,s_0)\, I_{2^- 2^-_c}(s,s_0)
\,\frac{b_{2^-_c}(s_0)}{b_{2^-}(s_0)}\hskip5ex 2^-\\
\alpha_1^{0c}(s,s_0)=\lambda+\alpha_1^1(s,s_0)\, I_{1^+_c 2^-}(s,s_0)
\,\frac{b_{2^-}(s_0)}{b_{1^+_c}(s_0)}
+\frac{2}{3}\,\alpha_1^{0c}(s,s_0)\, I_{1^+_c 1^+_c}(s,s_0)
\hskip13.3ex 1^+_c\\
\hskip10ex
+\frac{1}{3}\,\alpha_1^{1c}(s,s_0)\, I_{1^+_c 2^-_c}(s,s_0)
\,\frac{b_{2^-_c}(s_0)}{b_{1^+_c}(s_0)}\\
\alpha_1^{1c}(s,s_0)=\lambda+\alpha_1^1(s,s_0)\, I_{2^-_c 2^-}(s,s_0)
\,\frac{b_{2^-}(s_0)}{b_{2^-_c}(s_0)}
+\,\alpha_1^{0c}(s,s_0)\, I_{2^-_c 1^+_c}(s,s_0)
\,\frac{b_{1^+_c}(s_0)}{b_{2^-_c}(s_0)}\, . \hskip7ex 2^-_c\\
\end{array} \right.\eqno (A36)$$

\vskip2ex
{\bf The equations of ${\bf (1,2)}$ singlet.}
\vskip2ex

$\Lambda_c$ $\frac{3}{2} ^{-}$:

$$\left\{
\begin{array}{l}
\alpha_0^0(s,s_0)=\lambda+\frac{1}{2}\,\alpha_0^{0c}(s,s_0)
\, I_{0^+ 0^+_c}(s,s_0)\,\frac{b_{0^+_c}(s_0)}{b_{0^+}(s_0)}
+\frac{3}{2}\,\alpha_1^{1c}(s,s_0)\, I_{0^+ 2^-_c}(s,s_0)
\,\frac{b_{2^-_c}(s_0)}{b_{0^+}(s_0)}\hskip5ex 0^+\\
\alpha_0^{0c}(s,s_0)=\lambda+\alpha_0^0(s,s_0)\, I_{0^+_c 0^+}(s,s_0)
\,\frac{b_{0^+}(s_0)}{b_{0^+_c}(s_0)}
-\frac{1}{2}\,\alpha_0^{0c}(s,s_0)\, I_{0^+_c 0^+_c}(s,s_0)
\hskip5ex 0^+_c\\
\hskip10ex
+\frac{3}{2}\,\alpha_1^{1c}(s,s_0)\, I_{0^+_c 2^-_c}(s,s_0)
\,\frac{b_{2^-_c}(s_0)}{b_{0^+_c}(s_0)}\\
\alpha_1^{1c}(s,s_0)=\lambda+\alpha_0^0(s,s_0)\, I_{2^-_c 0^+}(s,s_0)
\,\frac{b_{0^+}(s_0)}{b_{2^-_c}(s_0)}
+\frac{1}{2}\,\alpha_0^{0c}(s,s_0)\, I_{2^-_c 0^+_c}(s,s_0)
\,\frac{b_{0^+_c}(s_0)}{b_{2^-_c}(s_0)}\hskip7ex 2^-_c\\
\hskip10ex
+\frac{1}{2}\,\alpha_1^{1c}(s,s_0)\, I_{2^-_c 2^-_c}(s,s_0)\, .\\
\end{array} \right.\eqno (A37)$$

}
\end{document}